\documentclass{scrartcl}

\usepackage{amssymb}
\usepackage{amsmath}
\usepackage[english]{babel}
\usepackage{booktabs}
\usepackage{hyperref}
\usepackage{graphicx}
\usepackage{authblk}
\usepackage[sorting=none]{biblatex}
\bibliography{MD_simulation_of_Nafion_Vivod}
\usepackage{multirow}

\begin{document}

    \title{Molecular dynamics simulations of Nafion thin films at a platinum catalyst surface: Correlating structure with charging behaviour}

    \author[1]{Dustin Vivod\thanks{Corresponding author: d.vivod@fz-juelich.de}}
    \author[1]{Binny A. Davis}
    \author[1]{Tobias Binninger}
    \author[1,2]{Michael Eikerling}

    \affil[1]{\small{\textit{Theory and Computation of Energy Materials (IET-3), Institute of Energy Technologies, Forschungszentrum J\"ulich GmbH, 52425 J\"ulich, Germany}}}
    \affil[2]{\small{\textit{Chair of Theory and Computation of Energy Materials, Faculty of Georesources and Materials Engineering, RWTH Aachen University, 52062 Aachen, Germany}}}
    \date{}
    \maketitle

    \begin{abstract}
        Electrocatalysis is greatly influenced by the local reaction environment, which is governed by the structure of the catalyst, the distribution of the electrolyte, and the local electric field. In catalytic systems comprised of complex molecular species like ionomers, the distribution of electrolyte can vary substantially, resulting in divers local reaction environments. In order to gain atom-scale insight into this micro-environment we construct a model system consisting of a platinum surface, varying levels of water, and a Nafion thin film and conduct molecular dynamics simulations. We employ a construction based on Voronoi tesselation to assemble a dense film of ionomer that fully covers the platinum substrate. An energy analysis reveals that water film configurations with thickness of less then 13\,\AA\ are stable. Simulations with charged platinum surfaces are analysed in view of electrostatic conditions and differential capacitance of the interface configuration. Trends observed in these properties can be interpreted in view of the crowding of hydronium ions or the Nafion film at the platinum surface. The presented workflow can be easily applied to investigate novel ionomers for use in PEMFCs.
    \end{abstract}

    \section{Introduction}
        \label{introduction}
        
        Electrochemical energy conversion devices such as fuel cells and electrolysers utilise nanostructured catalyst layers (CL) at which redox reactions take place to produce or consume oxygen and hydrogen gas, associated with the consumption or generation of electrical energy. To provide proton-conducting pathways to the surface of the elctrocatalyst at which the reactions proceed, and to improve the structural integrity of the CL, ionomer is added to it. In the specific case of polymer electrolyte fuel cells (PEFCs) the use of platinum as the catalyst and Nafion as the ionomer is well established and serves as the benchmark.\cite{kusoglu_a, van_cleve_t} However, due to the cost of platinum and the rising health concerns around per- and polyfluorinatedalkyl substances (PFAS), which Nafion is made up of, new materials are sought after.\cite{panieri,lohmann, podder, baker}
        
        The microscopic interfacial region where the electrocatalyst and the electrolyte come together defines the local reaction environment (LRE). As the name suggests it comprises distributions of properties that determine the rates of electocatalytical reactions in the immediate vicinity of the catalyst, at atom-scale distance. Relevant properties include the structure and orientation of water molecules, ion densitites, as well as electrostatic potenial or electric field. Small changes in any of these properties can induce pronounced variations in the rates of targeted reactions and degradation processes.\cite{wang_t, fan_r, zhou_y} The importance of the LRE for the electrocatalytic activity has been discussed extensively in the literature.\cite{chen_c, zhiyuan_x, zhu_x} It is evident that in order to find replacements for catalyst and ionomer that yield required cost reductions, deliver gains in performance and stability, and alleviate environmental concerns, it is crucial to understand the LRE.
        
        In the nano-confined region of the LRE, the properties of the electrolyte differ dramatically from those of the bulk due to the close proximity of the catalyst. As such the interface region that defines the LRE can exhibit a range of structures. One such structural motif that is being discussed in the literature is the presence and thickness of a thin water film between electrode and ionomer.\cite{wood_d, chabot_a, amin} Recent studies by Chabot et al.\ used small-angle neutron scattering (SANS) to elucidate the structure of ionomer-bound interfaces in cathode catalyst layers of PEFCs, and thus provide valuable insight into the LRE.\cite{chabot_a, chabot_b} They interpreted SANS data with the presence of a thin water film between a skin-type ionomer layer and the catalyst. The water film thickness they observed was in the range of 7--13\,\AA, while the ionomer layer was between 20--30\,\AA\ thick. These thicknesses varied depending on the relative humidity and the carbon support material that was used. The structural motif of this thin-film morphology is further supported by compelling intuitive arguments. It underpins the need for well-devised models that adequately represent the interface structure and conditions, and form a reliable basis for analyses of transport phenomena and reaction kinetics at the nanoscale. 
        
        Other works, including modeling and simulation studies, investigated the effect of a water film and its thickness on the LRE.\cite{amin, victor_a, vuppala, jiang_s, min} These type of studies are vital, as experimental insights are difficult to obtain due to the nanoscale and complex nature of the LRE. A key benefit of these computational works is the possibility to investigate the influences of specific changes to a certain parameter such as water film thickness, ionomer structure or catalyst surface charge. Molecular dynamics (MD) simulations that investigate these influences have been carried out in studies that utilised a flat catalyst surface and a thin Nafion film that confine a water film of variable thickness. In simulations of Nouri-Khorasani et al., the concentration of protons located at the platinum surface was found to decrease with the increase in water film thickness.\cite{amin} Fernandez-Alvarez et al. on the other hand observed in their MD study that oxygen diffusion remains largely unaffected by a change in the water film thickness.\cite{victor_a} Both these MD studies provide atom-scale insight into spatial distributions of reactants. A key parameter that was evaluated in these studies is the charging state of the catalyst. As the surface charge at the catalyst varies with electrode potential during device operation, the effect of this charging has to be accounted for. The simplest way to achieve this is by adding charges \textit{ad hoc} to the platinum atoms that remain fixed over the course of the simulation, corresponding to the fixed charged method (FCM). The more complex constant potential method (CPM), on the other hand, evaluates the point charges of a selected group of atoms at each time step at a set potential.\cite{siepmann, reed} The CPM adds computational complexity with the argued benefit of being physically consistent. Zhang et al.\ compared the results for simulations of both FCM and the CPM using the otherwise same setup of a Nafion layer on a platinum surface.\cite{zhang} However, with their setup they saw no change in the resulting electrostatic potential distribution. While the charging state was considered in these studies, the resulting electrostatic properties were not investigated in full detail, leaving vital insight unreported.

        Other MD studies exist that have simulated the LRE in more complex systems of platinum nanoparticles supported on carbon with the Nafion phases created from annealing in order to bridge the gap to more realistic structures found in PEFCs.\cite{victor_b, garcia_2021, xue_2022, kim_d, you_2024, garcia_2024} However due to the increase in complexity of these systems the influence of specific parameters on the LRE is difficult to discern, thus they are more focused on investigating the transport properties of hydronium ions and the influence of the Nafion composition. 
        For a more extensive review on the state of MD simulations of PEFCs we refer to a recent publication by Rayhani and Jian.\cite{rayhani}
        A physically coherent picture of the interplay of the interfacial water film with the properties of the ionomer film and the charging state of the catalyst jointly determining the electrostatic properties, is however still missing.

        In this study we utilise a thin-film structural model of an atom-scale thin slab of water that is confined between a planar platinum (111) surface and a quasi 2D skin layer of Nafion, thus resembling the layered structure identified by Chabot et al.\cite{chabot_a, chabot_b} Our thin-film model uses freely moving strands of Nafion and is set-up with the help of Voronoi tesselation. Multiple systems are created that vary in water content to explore the effect of the water slab thickness. Changes to the charging state of the platinum surface are applied using the FCM. In this publication, we neglect the impact of the potential-dependent formation of chemisorbed oxygen species. The fixed charges imposed at the catalyst surface and at the ionomer skin are neutralized by adjusting the number of hydronium ions included in the simulations. The system is then investigated by complementary structural and electrostatic analyses. The structural analyses focus on providing an atom-scale picture of the LRE with varying water film thickness. We demonstrate how to obtain the differential capacitance we that deem of practical utility as a fingerprint of interface structured LRE.

    \section{Computational Details}
      MD simulations reported in this article have been performed using the LAMMPS \cite{lammps} code for classical molecular dynamics simulation with force field (FF) parameters taken from previous publications.\cite{mabuchi,brunello} A cutoff of 12\,\AA\ was used for all interatomic potentials as well as for the direct part of the electrostatic interactions. Electrostatic interactions were calculated using the particle--particle--particle--mesh method with the slab correction and an accuracy of 10$^{-5}$\,eV\,\AA$^{-1}$.\cite{hockney} Periodic boundary conditions were employed in X and Y direction.
      A reflective wall has been added to all simulations at a distance of 60\,\AA\ away from the platinum surface. Any atom that would cross this wall is reflected back towards the platinum surface by flipping the sign of the velocity in Z direction. Simulations were carried out at temperatures of 298.15\,K, i.e.\ room temperature (RT), and 353.15\,K, controlled by a Nos\'e-Hoover thermostat with a damping constant of 0.5\,ps. The time step for all simulations was 1\,fs.
      
      The platinum slab used was cut from the bulk crystal with a lattice constant of 3.94\,\AA. The surface of the slab had (111) orientation and the slab consisted of 9 layers. During the simulations the bottom 3 layers of the slab were frozen. The dimensions of the slab were 99.11 $\times$ 100.1 $\times$ 18.93\,\AA\textsuperscript{3}. The complete simulation cell consisted of the platinum slab on one side and a constraining Nafion layer on the other side with the space in between filled with varying amounts of water molecules, namely 0, 1000, 5000, and 9000 water molecules. In the first step the systems with only platinum and water were constructed and equilibrated for 6\,ns at both RT and 353.15\,K under NVT conditions. The resulting approximate water-film thicknesses are listed in Tab.\ \ref{tab:wft}
      
      \begin{table}
        \centering
        \caption{\label{tab:wft}Approximate water film thicknesses in simulations of platinum-water systems at 298.15\,K.}
        \begin{tabular}{cc}
            \toprule
            \# Water molecules & Film thickness [\AA] \\ \midrule
            1000& 1.5\\
            5000& 13\\
            9000& 26\\ \bottomrule
        \end{tabular}
    \end{table}

      \subsection{Nafion layer setup}

      \begin{figure}[h!]
          \centering
          \includegraphics[width=.25\textwidth]{figure_1.png}
          \caption{\label{fig:naf_struct}Chemical structure of Nafion with backbone and side-chain parts highlighted in green and orange, respectively. In this study n=7 and m=5.}
      \end{figure}

\begin{figure}[!ht]
        \centering
        \includegraphics[width=.3\textwidth]{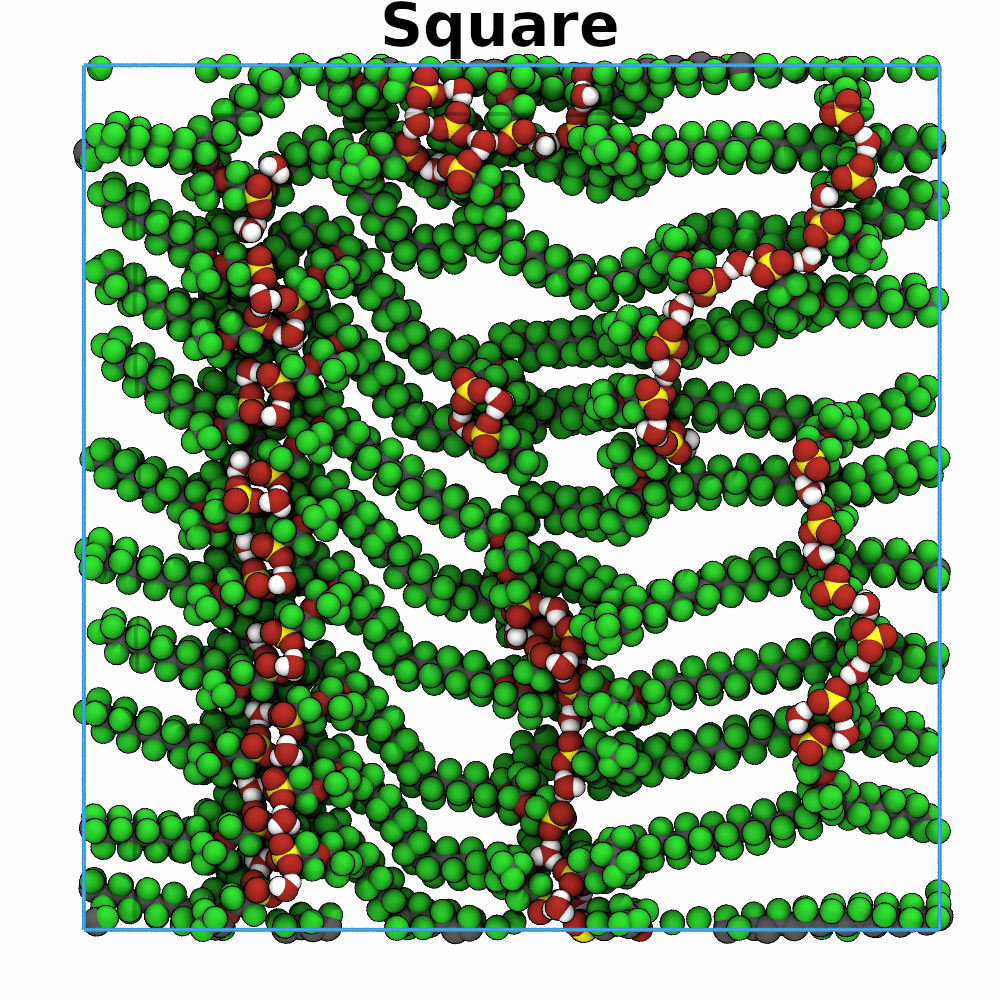}
        \includegraphics[width=.3\textwidth]{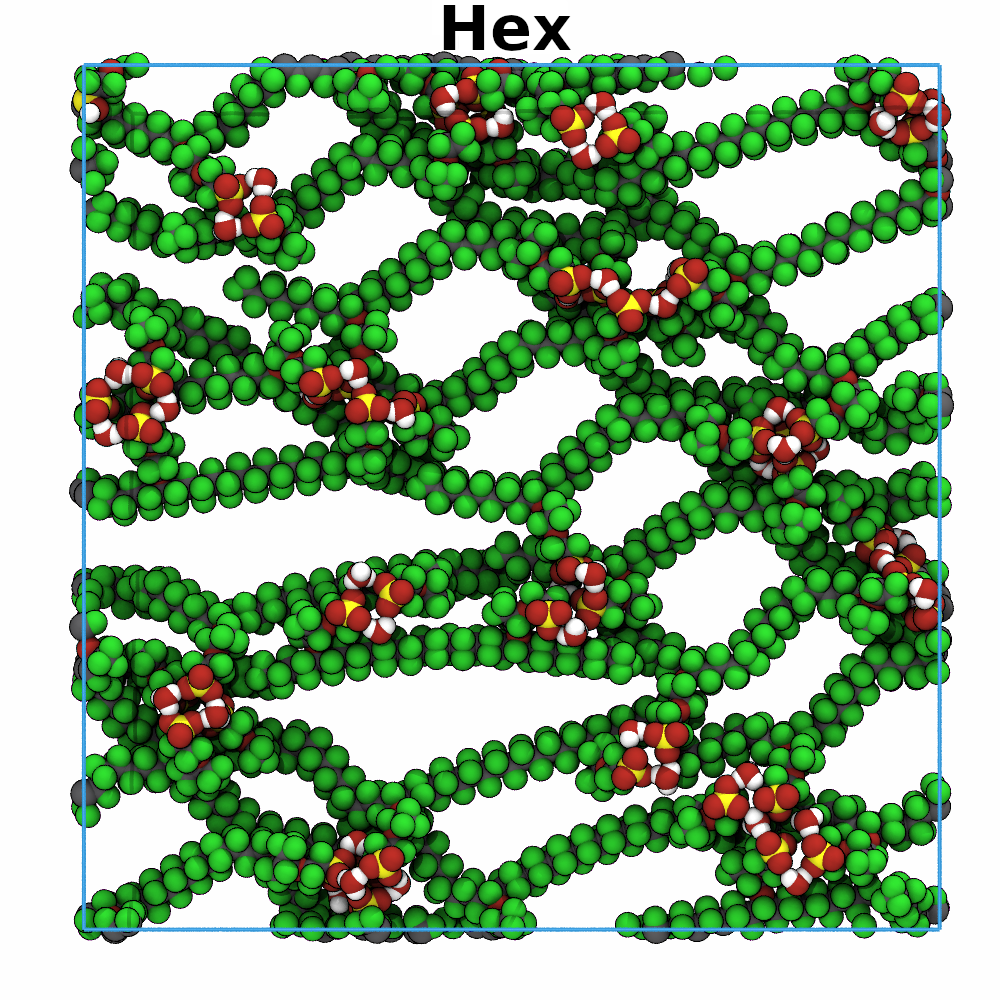}
            \caption{\label{fig:naf_layers}Exemplary snapshots of the two initial Nafion layers used within this work. Hydronium ions were added for charge neutrality. Color code: H: white, C: black, O: red, F: green, S: yellow.}
        \end{figure}

      To construct a dense Nafion film that covers the entire platinum slab, we proceed by first constructing auxiliary thin-film models of low density. To this end, a Nafion strand was created with five monomer repeat units (cf.\ Fig.\ \ref{fig:naf_struct}). The oligomer strand has an equivalent weight of 1122\,g\,mol\textsuperscript{-1}, which  closely resembles the chemical composition of Nafion 117 as the most commonly variant used in PEFCs.\cite{jang} The strand was then stretched in X direction to fit the dimensions of the Pt slab used. The stretched strand is then replicated to construct the thin film. With this strand two different thin-film types were assembled. For the first type denoted as `square' the strand was replicated in Y direction with a shifting vector of length 10\,\AA. The other, denoted as `hex' was created by shifting of the strand in Y and X by 10\,\AA\ in both directions. The resulting layers, each consisting of 10 strands, are shown in Figure~\ref{fig:naf_layers}. The number of strands was chosen such that the initial strands do not overlap. All sulfonic acid head groups were deprotonated, and as a result the Nafion molecules carried a negative charge. To ensure charge neutrality, the corresponding number of hydronium ions (H$_3$O$^+$) was added to the simulation cell. While the model of fully deprotonated Nafion in an environment with no water molecules is not physically motivated, we use these systems for comparison. By using deprotonated Nafion and hydronium ions no change in the FF is necessary when going to hydrated systems thus avoiding the introduction of errors due to FF inconsistencies.

      These layers were added atop the platinum-water systems created before. The combined systems were simulated for a minimum of 5\,ns. To verify that equilibration had been achieved, the total energy was monitored over the last 5\,ns of simulation time. The data set was split into two halves, and a histogram was constructed for the energies of each set. The histograms were fitted via Gaussians, and equilibration was considered to be achieved when the difference between the two peak centers was less than 1\% from one another. After this equilibration, production runs of up to 40\,ns of total simulation time were performed. The total simulation time needed was determined by the contacts of hydronium ions with the platinum surface (cf. SI\ Fig.\ \ref{fig:contacts}), with simulation times being increased as long as an increase in contacts was observed.
      
        \begin{figure}[!ht]
            \centering
            \includegraphics[width=.3\textwidth]{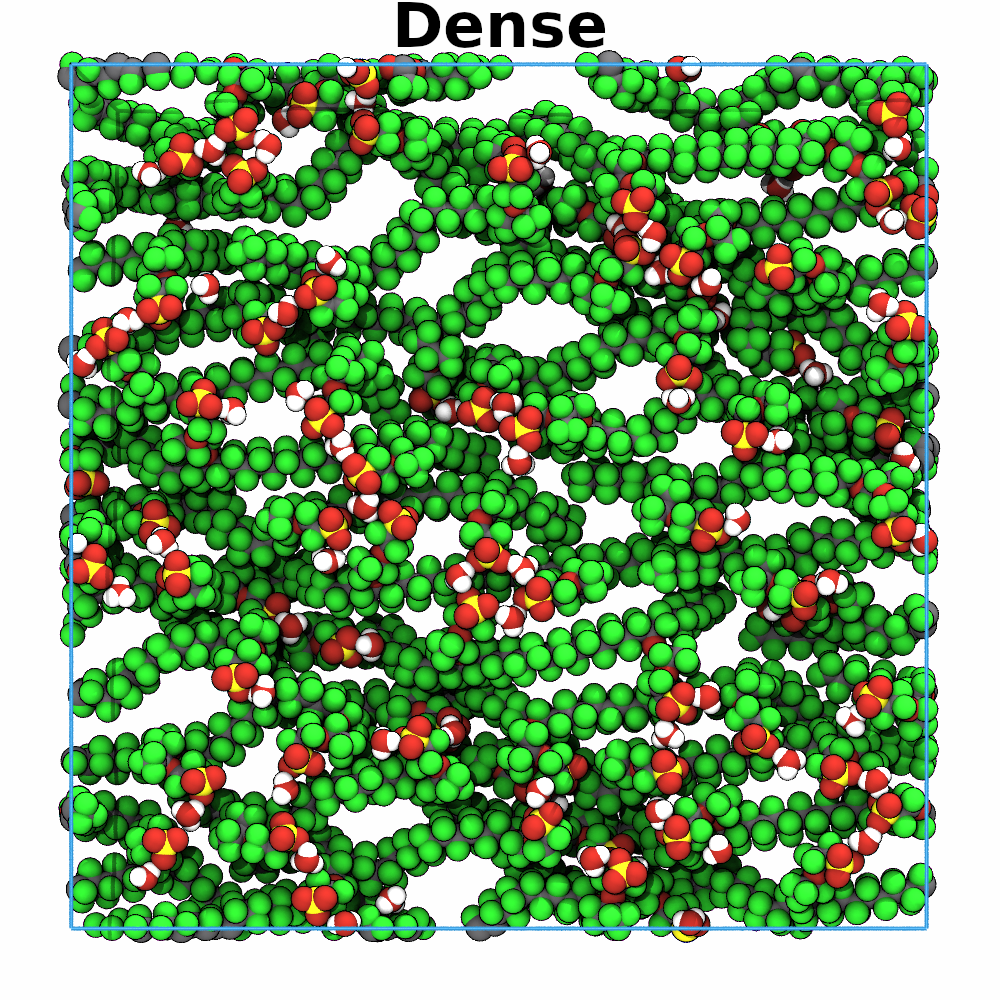}
            \caption{\label{fig:naf_fl_layer}Exemplary snapshot of the dense Nafion layer. Hydronium ions were added for charge neutrality. Color code: H: white, C: black, O: red, F: green, S: yellow.}
        \end{figure}

\begin{figure}[!ht]
            \centering
            \begin{tabular}{cc}
                (a) & (b)\\
                \includegraphics[width=.24\textwidth]{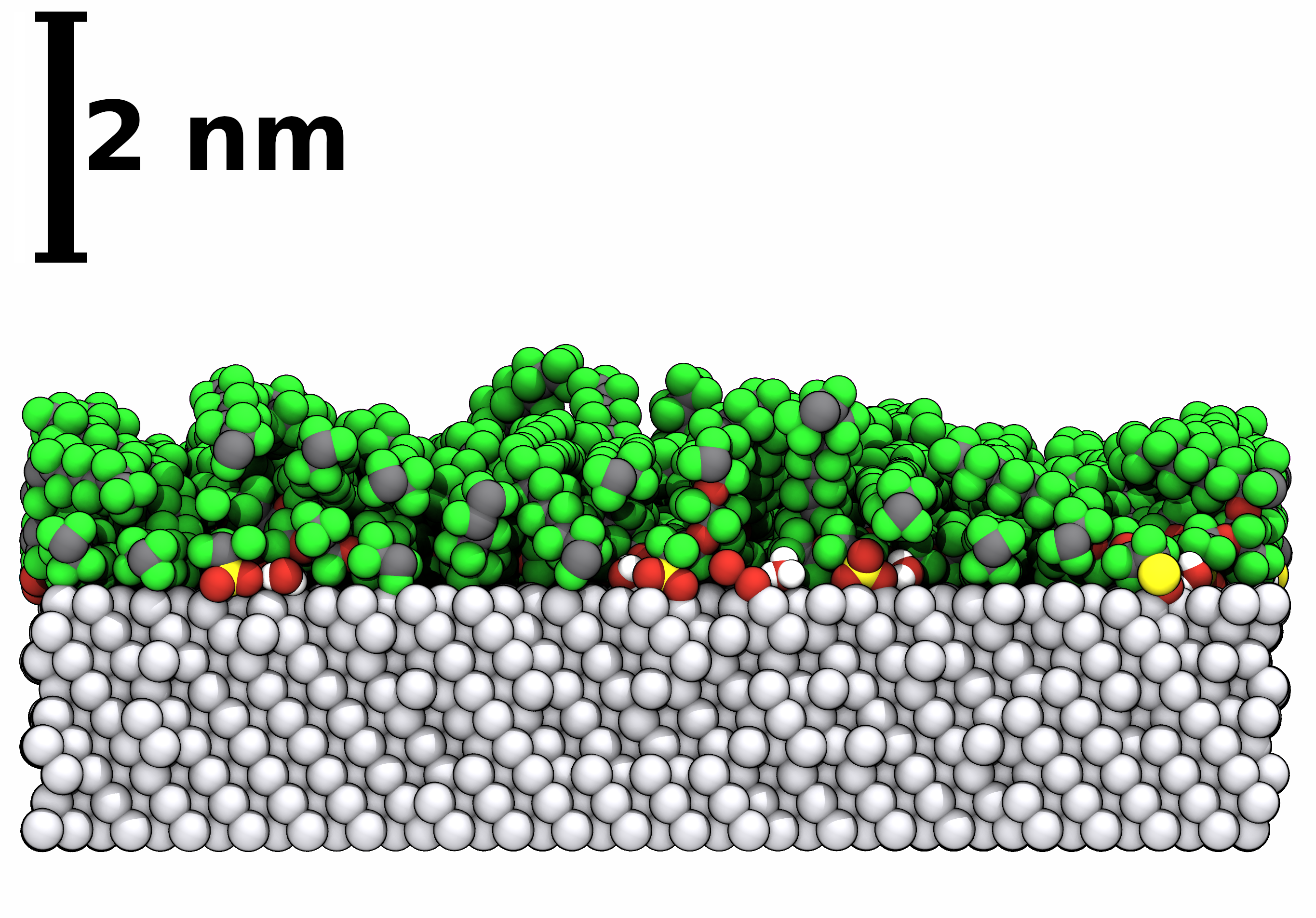} & \includegraphics[width=.24\textwidth]{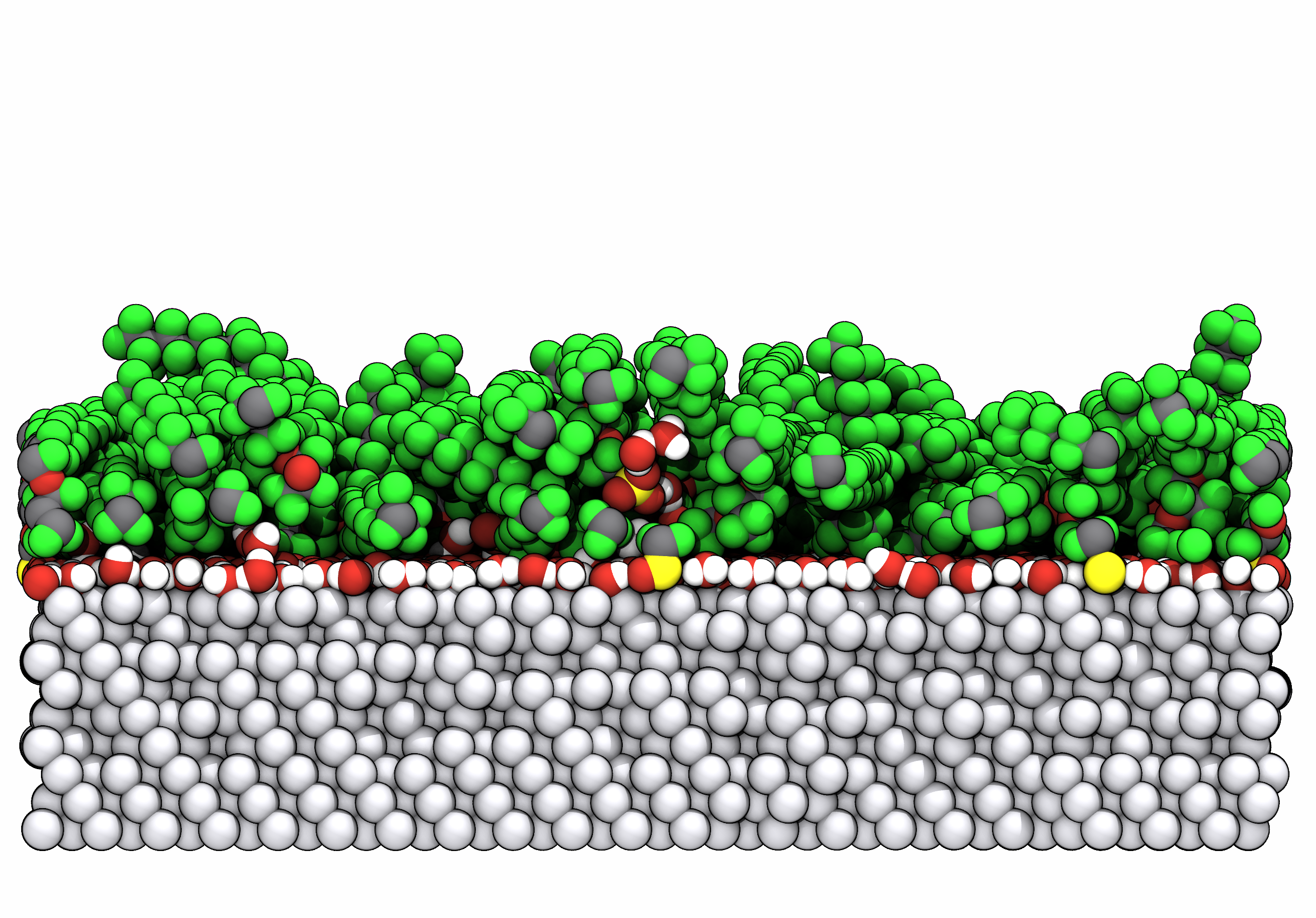}\\
                (c) & (d)\\
                \includegraphics[width=.24\textwidth]{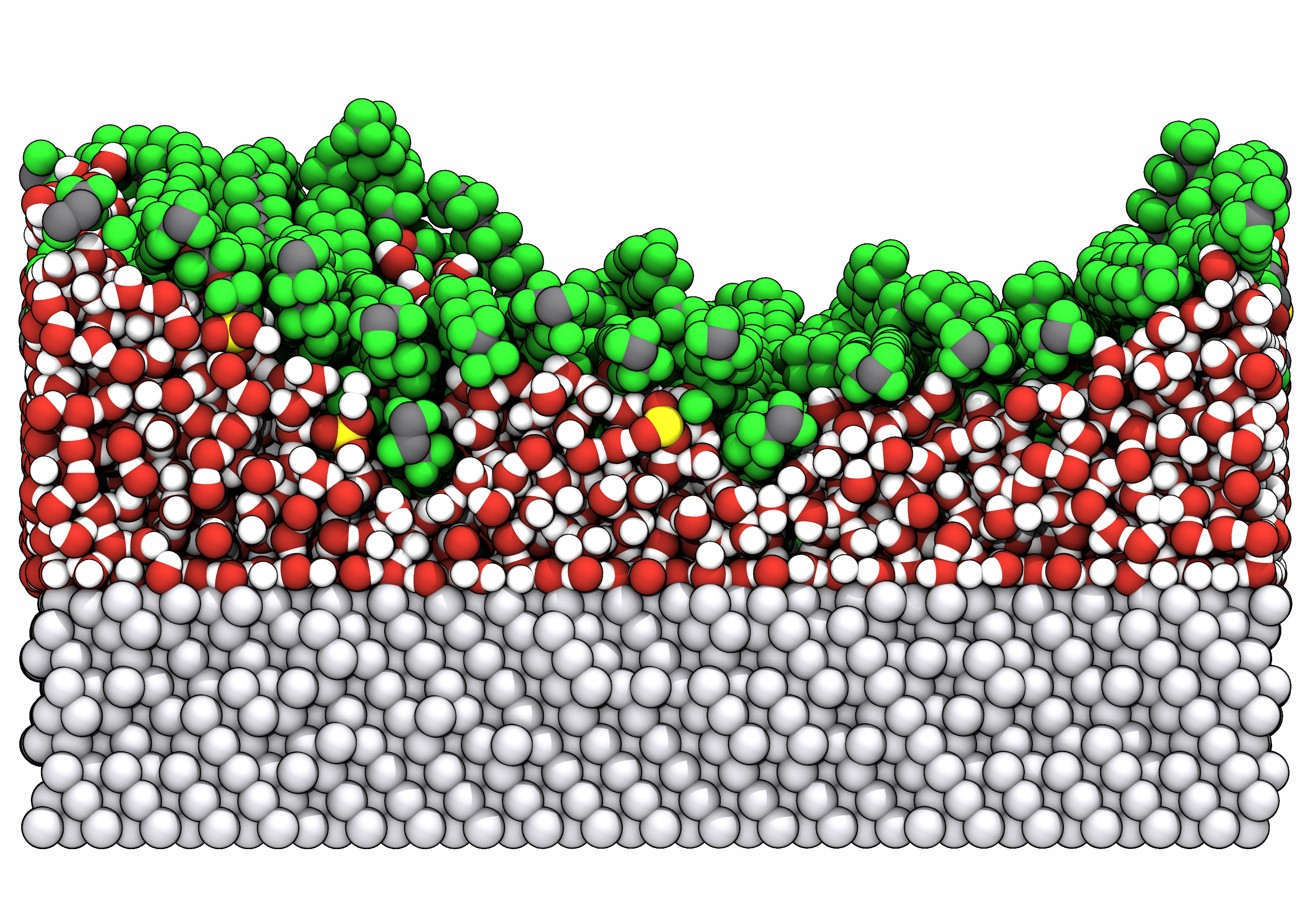} & \includegraphics[width=.24\textwidth]{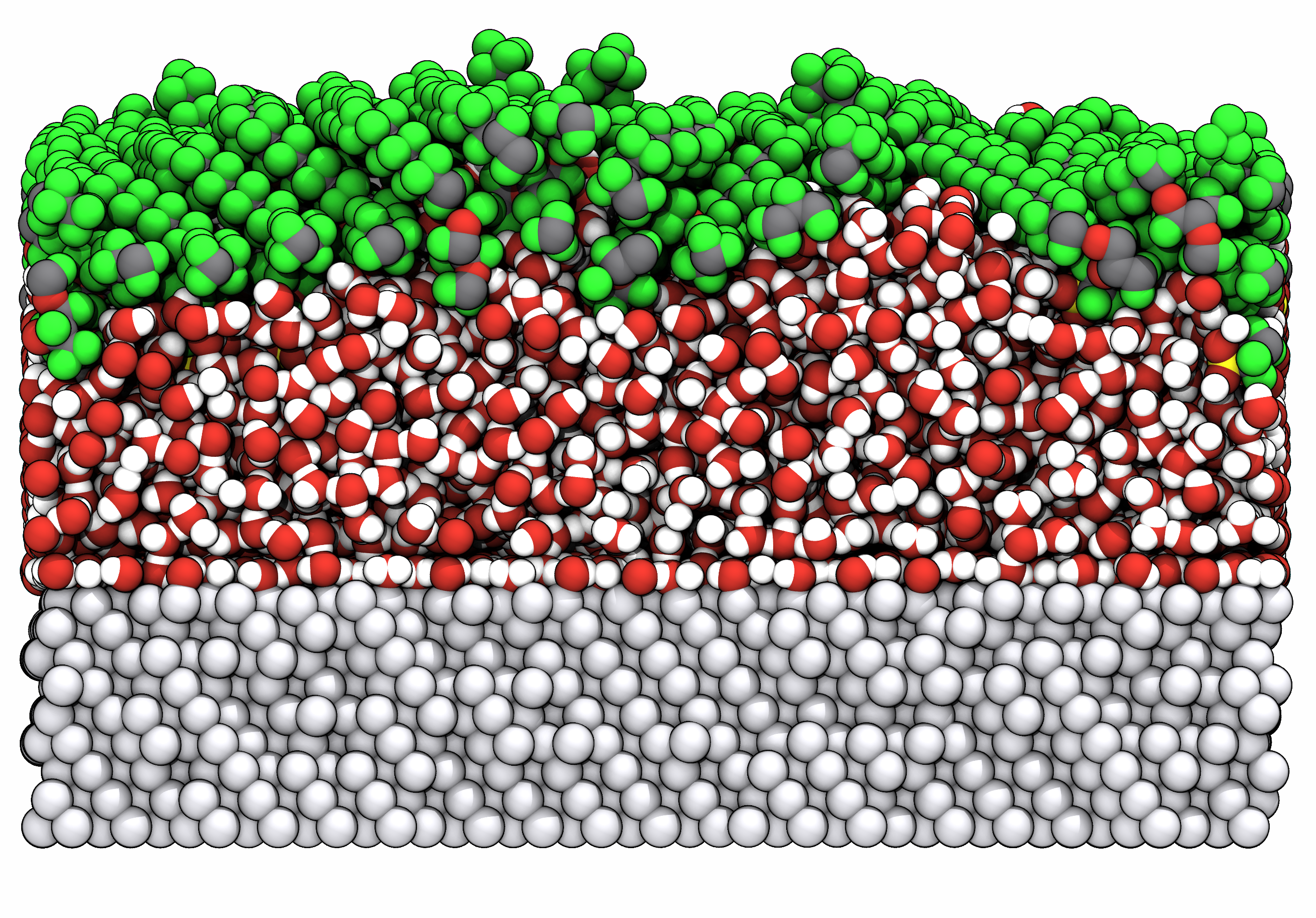}
            \end{tabular}
            \caption{\label{fig:snaps}Snapshots during production runs of systems with different water content. (a): 0 water molecules, (b): 1000 water molecules, (c): 5000 water molecules, (d): 9000 water molecules. Color code: H: white, C: black, O: red, F: green, S: yellow, Pt: silver.}
        \end{figure}

\begin{figure}[!ht]
        \centering
        \includegraphics[width=\textwidth]{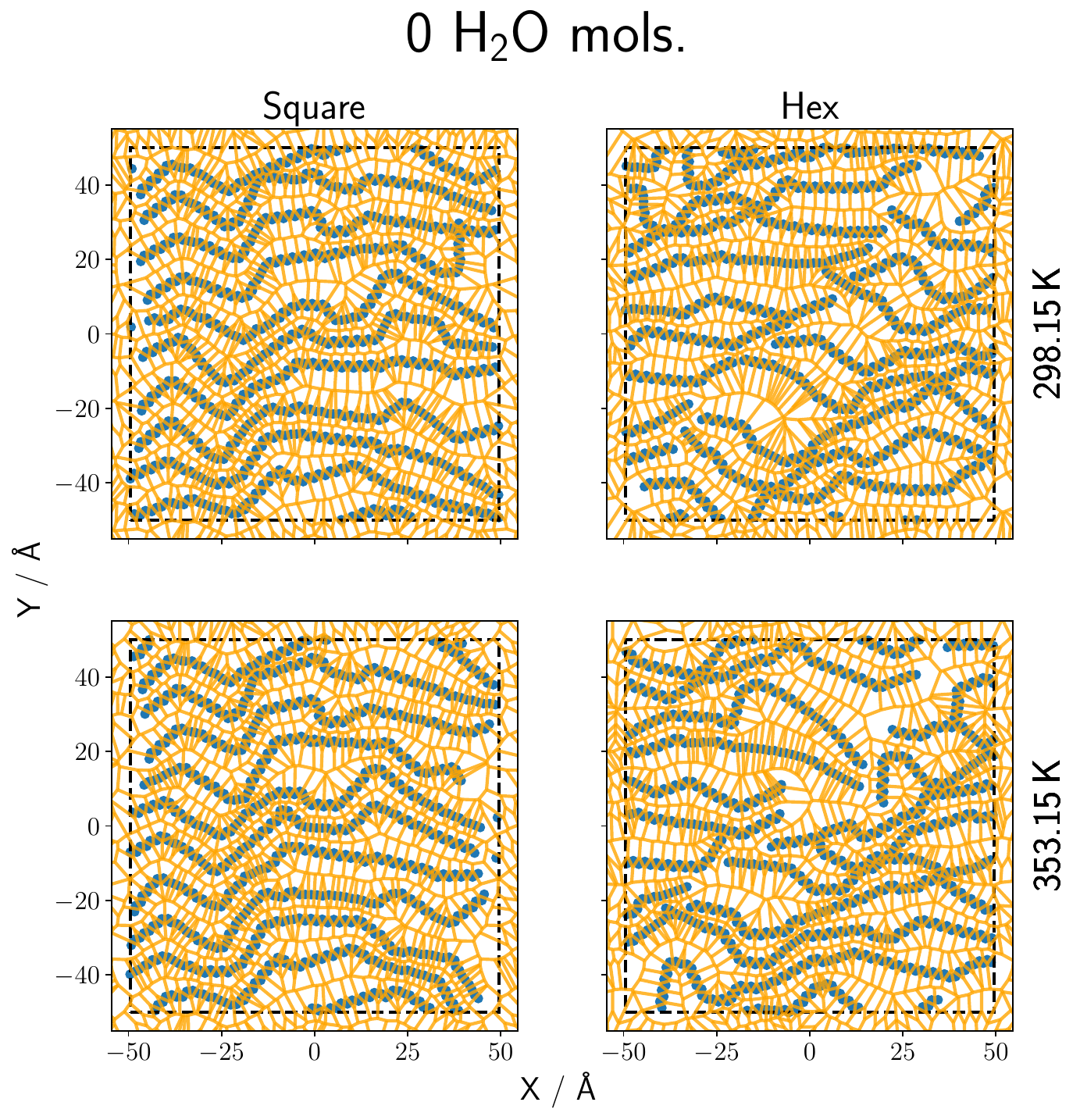}
        \caption{\label{fig:voronoi_0}Exemplary Voronoi diagrams for the square and hex Nafion layers (column wise) at both investigated temperatures (row wise) and 0 added water molecules. Blue dots denote carbon backbone atoms, orange lines denote the region edges. Black dashed lines show the simulation cell boundaries.}
    \end{figure}

\begin{figure}[!ht]
        \centering
        \includegraphics[width=\textwidth]{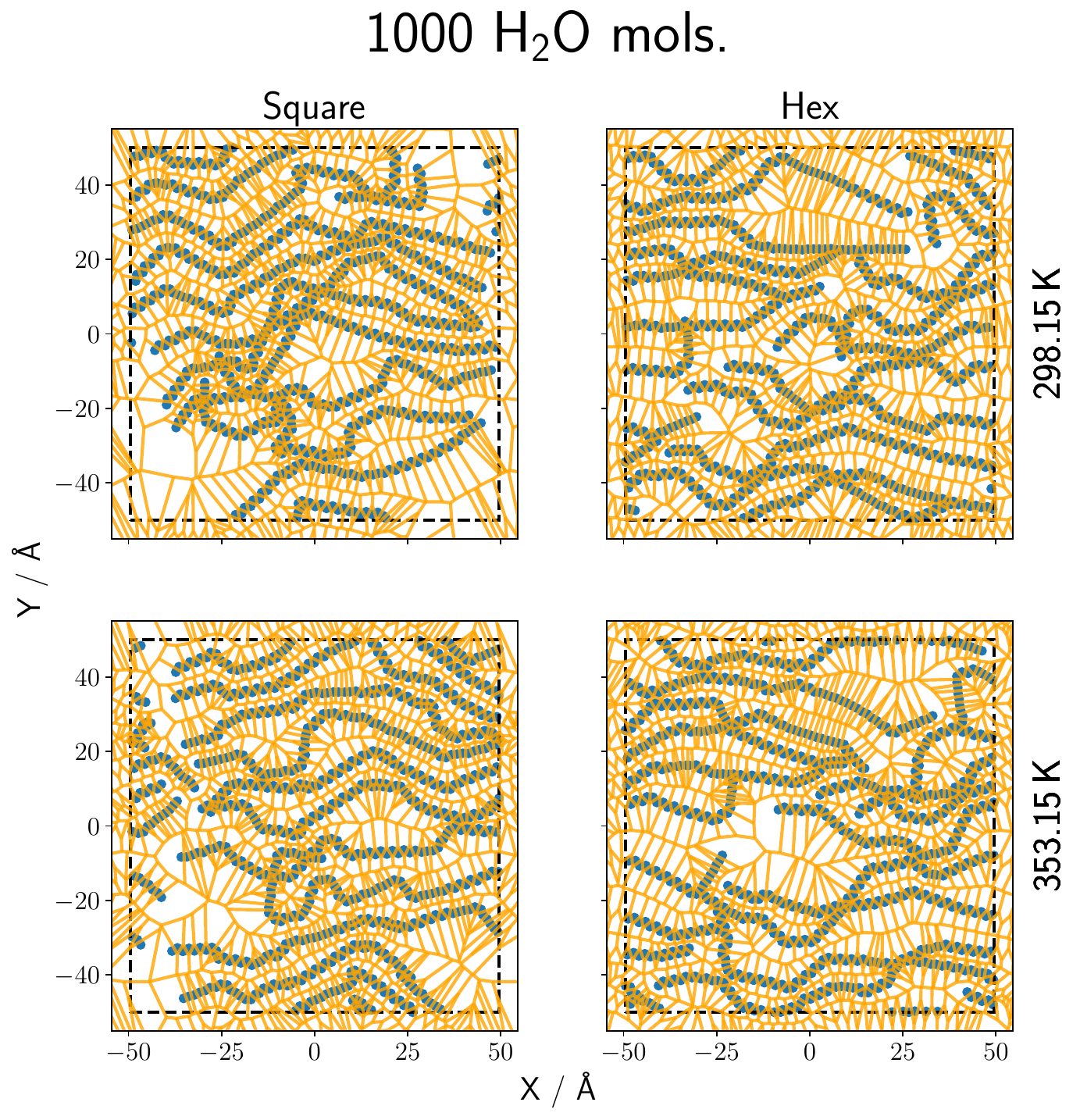}
        \caption{\label{fig:voronoi_1000}Exemplary Voronoi diagrams for the square and hex Nafion layers (column wise) at both investigated temperatures (row wise) and 1000 added water molecules. Blue dots denote carbon backbone atoms, orange lines denote the region edges. Black dashed lines show the simulation cell boundaries.}
    \end{figure}

\begin{figure}[!ht]
        \centering
        \includegraphics[width=\textwidth]{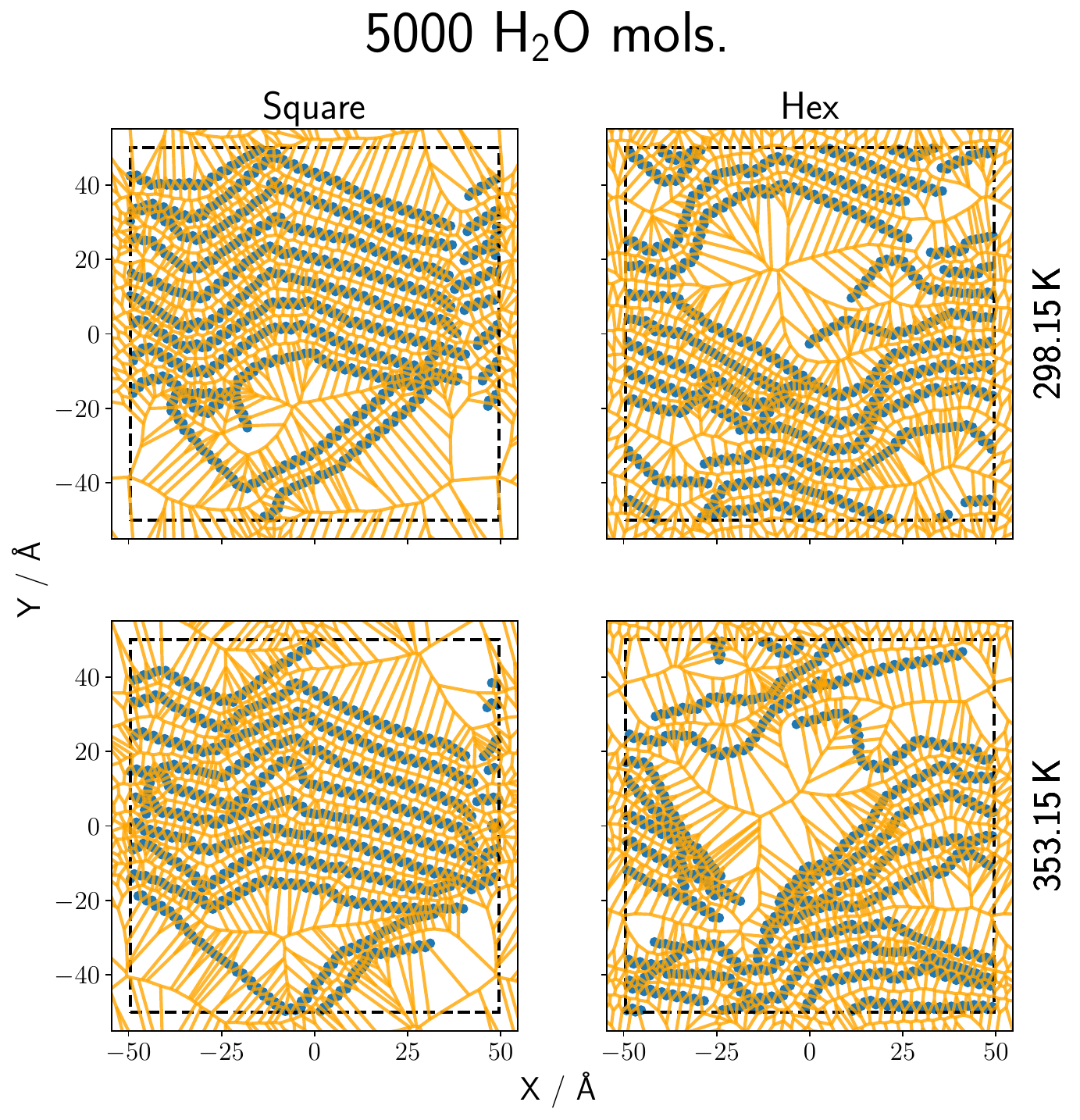}
        \caption{\label{fig:voronoi_5000}Exemplary Voronoi diagrams for the square and hex Nafion layers (column wise) at both investigated temperatures (row wise) and 5000 added water molecules. Blue dots denote carbon backbone atoms, orange lines denote the region edges. Black dashed lines show the simulation cell boundaries.}
    \end{figure}

\begin{figure}[!ht]
        \centering
        \includegraphics[width=\textwidth]{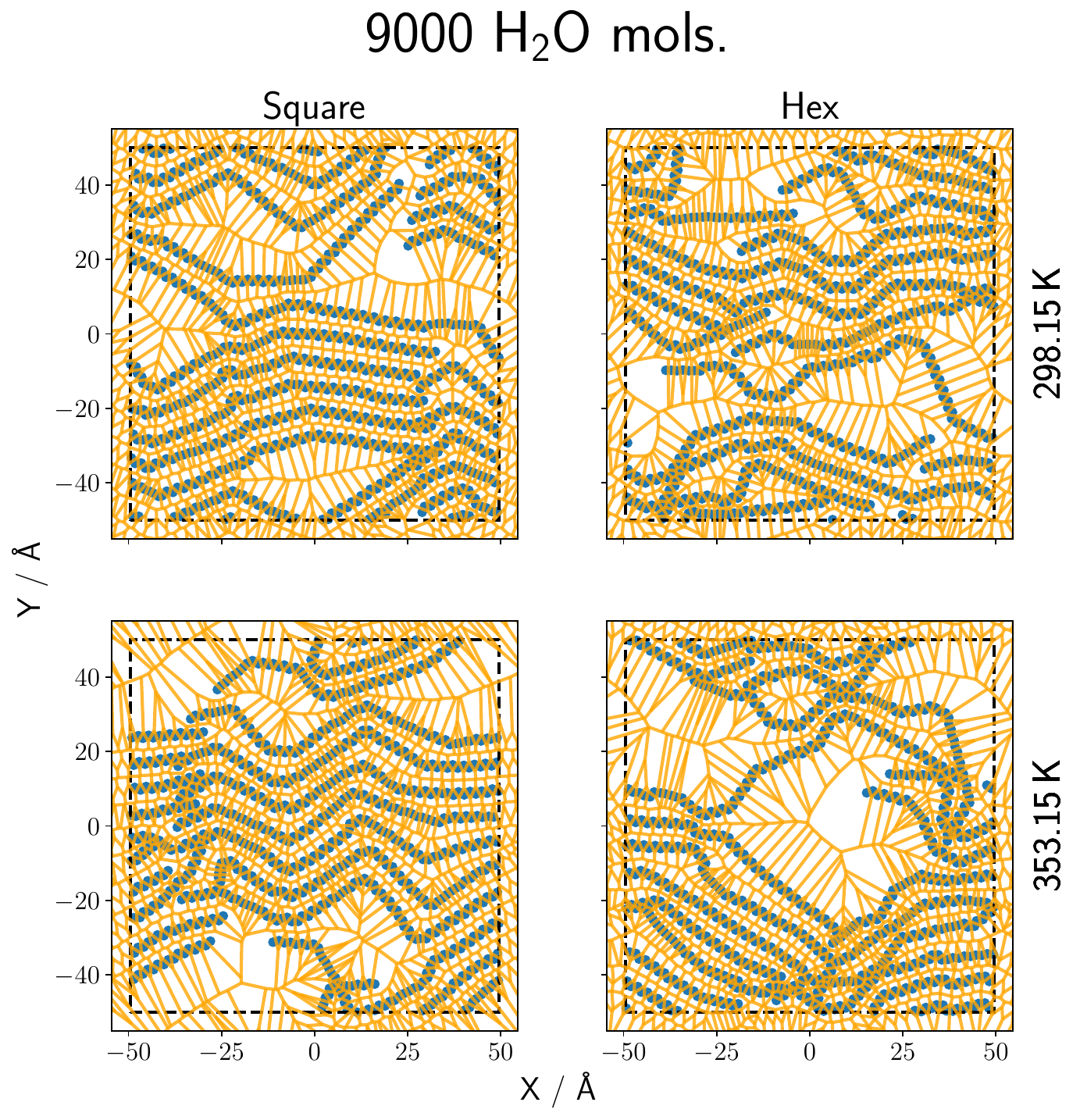}
        \caption{\label{fig:voronoi_9000}Exemplary Voronoi diagrams for the square and hex Nafion layers (column wise) at both investigated temperatures (row wise) and 9000 added water molecules. Blue dots denote carbon backbone atoms, orange lines denote the region edges. Black dashed lines show the simulation cell boundaries.}
    \end{figure}

        Afterwards, we performed Voronoi tesselations of the square and hex Nafion layers of all interface systems to rationalise a dense thin film of Nafion that uniformly covers the platinum slab. The Voronoi tesselation as implemented in SciPy generates a diagram of areas surrounding a set of given data points in 2D space.\cite{scipy} Each point within a given area is closest to the data point, around which the area was constructed. By taking the X and Y coordinates of the Nafion backbone carbon atoms as data points we interpret the area surrounding each data point as the spatial demand of the atom. As the Z coordinate is neglected, it has to be noted that errors are introduced in cases where backbone atoms overlap. 
        For each trajectory, the Voronoi tesselation was performed at each frame, which are written every 1\,ps, (cf.\ SI\ Fig.\ \ref{fig:voronoi_0}-\ref{fig:voronoi_9000}), and the minimum, maximum and median area sizes are noted (cf.\ SI\ Tab.\ \ref{tab:voronoi_si}). With the mean median area sizes collected from all trajectories, we estimate the number of Nafion strands necessary to cover the platinum slab. Dividing the area of the platinum slab (99.2\,nm$^2$) by the mean median area of a single carbon backbone atom (10.6\,\AA$^2$) gives us the number of backbone carbon atoms that would fit onto the surface, within reasonable error. The number of backbone atoms is related to the number of Nafion strands via the equivalent weight. By using the same equivalent weight as before, i.e., 1122\,g\,mol$^{-1}$, our simulation cell required 14 Nafion strands to completely cover the platinum surface. The resulting Nafion layer, depicted in Fig.\ \ref{fig:naf_fl_layer}, is called `dense' and constructed by shifting the original strand by 10\,\AA\ in X and 7.5\,\AA\ in Y. The layer was then used to construct eight simulation cells with the platinum slab described above and 0, 1000, 5000, and 9000 added water molecules (cf. Fig.\ \ref{fig:snaps}) at RT and 353.15\,K with the same equilibration and production run protocol as detailed above. All further discussed simulations used this skin layer.

        In order to compare the total energies of the platinum-water-Nafion systems, two additional simulations of pure water were carried out, one for each temperature considered here. These water systems contained 12000 water molecules and were exposed to periodic boundary conditions in all cardinal directions. The simulations were performed under NPT conditions using the Nos\'e-Hoover barostat with a damping constant of 5\,ps and a target pressure of 1.01325\,bar. They were equilibrated for 5\,ns after which production runs of 5\,ns were carried out, of which the energies where sampled.

      \subsection{Electrostatic potential and differential capacitance}
        \label{sec:esp_dl}
        By removing or adding hydronium ions to the system, a charge imbalance was created which was compensated by tuning the total charge of the surface Pt atoms (e.g., from 0 for the initial systems to +10\,$e$, corresponding to $\sim$0.007\,$e$ per surface platinum atom, for systems with 10 hydronium ions removed from the electrolyte slab).

        In order to evaluate the capacitance of the interface, we first need to calculate the electrostatic interface potential. We start, analogous to Lin et al.\cite{lin}, from the Poisson equation,
        \begin{equation}
            \label{eq:poisson}
            \nabla^2 \phi = -\frac{\rho}{\epsilon_0},
        \end{equation}
        where $\phi$ is the potential, $\rho$ is the total charge density (including electrolyte species, vis. including solvent ions, as well as polymer constituents), and $\epsilon_0$ is the vacuum permittivity. Since our system is 2D periodic in X and Y, and the interface spans along Z, we average the charge density in the X-Y-plane and rewrite Eq.\ \ref{eq:poisson} as
        \begin{equation}
            \label{eq:poisson_z}
            \frac{d^2 \phi(z)}{dz^2} = -\frac{\rho(z)}{\epsilon_0},
        \end{equation}
        where $\rho(z)$ is the average charge density in the X-Y-plane and $\phi(z)$ the electrostatic potential at position $z$. Integrating Eq.\ \ref{eq:poisson_z} leads to
        \begin{equation}
            \label{eq:poisson_z_int}
            \phi{(z)} = -\frac{1}{\epsilon_0} \int \int \rho{(z)} dz du 
        \end{equation}

        The two emerging integration constants are fixed by setting the potential and electric field to zero at z = $\infty$, which in our case corresponds to the edge of the simulation box at the maximum Z value $Z_{max}$. Thus, the potential is integrated from the vacuum side, starting from a value of 0, going through the Nafion layer, water film, and finally the platinum slab (cf. Fig~\ref{fig:snaps}), yielding the upper integration bound of the minimum Z value $Z_{min}$ of the simulation box. We evaluate the potential with respect to a vacuum reference and not with respect to an electrode reference as is oftentimes found in literature.\cite{zhang, lin} While this choice has no effect on the numeric results, it offers better insight from plots of the potential distribution. 
        The difference in electrostatic potential between the platinum surface and the reference point is the surface potential $\phi_s$, which is equivalent to the electrode potential. The electrode potential of the uncharged simulations can then be related to the experimentally obtained electrode potential at the point of zero charge (PZC) found in literature. For a platinum (111) electrode the PZC is $\sim$0.23\,V vs. SHE.\cite{xu_p}

\begin{figure*}[!ht]
        \centering
        \includegraphics[width=\textwidth]{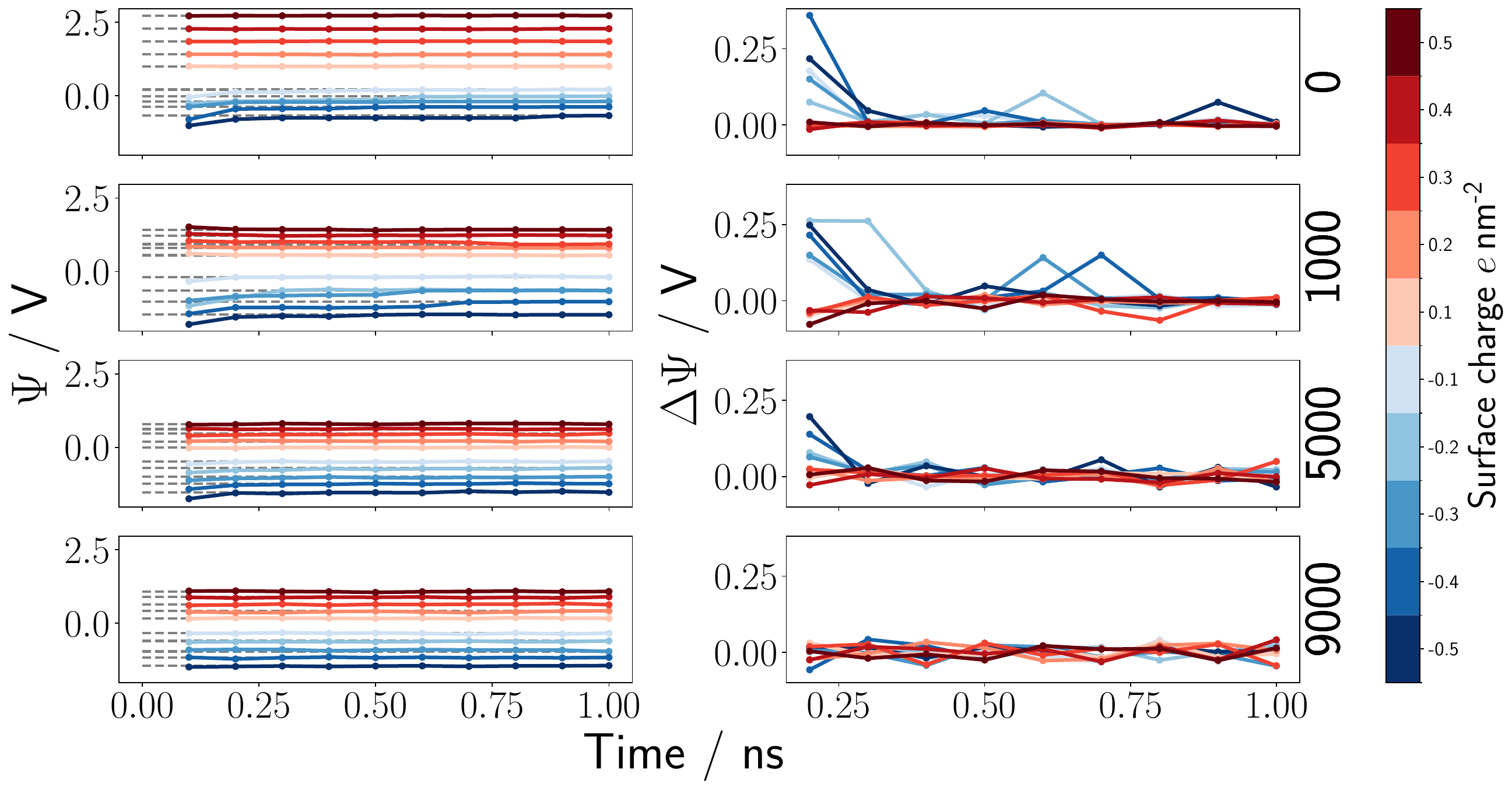}
        \caption{\label{fig:pot_time_evo}Left: Temporal evolution of the potentials at the platinum surface during the 1\,ns simulation. Gray dashed lines denote the averaged potential of the last 100\,ps. Right: Difference in the potential with the previous 100\,ps.}
    \end{figure*}

        The Pt surface was charged uniformly by fixed charges. Zhang et al.\ compared the difference of the fixed charge method (FCM) with the constant potential method (CPM) for a similar Pt-Nafion interface, where the Nafion was randomly dispersed and annealed atop the platinum surface.\cite{zhang} In their evaluation, both methods led to similar results, with the FCM being much less computationally intensive. Thus we also use the FCM in our study.
        We consider 10 different surface charges ranging from $-$0.5 to 0.5\,$e$\,nm$^{-2}$ with a step size of 0.1\,$e$\,nm$^{-2}$. The extra charges are balanced by adding or removing hydronium ions, where 10 hydronium ions correspond to a charge of $\sim$0.1\,$e$\,nm$^{-2}$, thus changing the topology of the system. To evaluate the effect of this change, each system was propagated for 1\,ns during which the electric surface potential was calculated by averaging over windows of 100\,ps. With this, we could verify that the potentials were equilibrated within $\sim$0.5\,ns (cf.\ SI\ Fig.\ \ref{fig:pot_time_evo}). We thus used the last 0.5\,ns for analysis of the averaged charge densities and the potentials. Each system was set up 3 times using different random number seeds for hydronium insertion or deletion, and the average charge densities are used for analysis.

        The differential capacitance was calculated via, the following relation
        \begin{equation}
            \label{eq:dlcap}
            C_{\mathrm{diff}} = \frac{\Delta\sigma}{\Delta\phi_s},
        \end{equation}
        where $\Delta\sigma$ is the change in surface charge between adjacent simulated charge states (in our case it is a constant value of 0.1\,$e$\,nm$^{-2}$), and $\Delta\Psi$ is the corresponding change in electric surface potential. The reported differential capacitances are calculated by considering the increase and decrease in surface charge for each data point and taking the average.

        Structural analyses were carried out using MDAnalysis and VMD was used to render the figures of the simulation cells.\cite{mda1, mda2, vmd1, vmd2} Unless stated otherwise, only the last nanosecond of each trajectory was used for analysis.

    \section{Results and Discussion}
        \subsection{Force field validation}
            \label{sec:validation}
\begin{figure}[!ht]
                \centering
                \includegraphics[width=.4\textwidth]{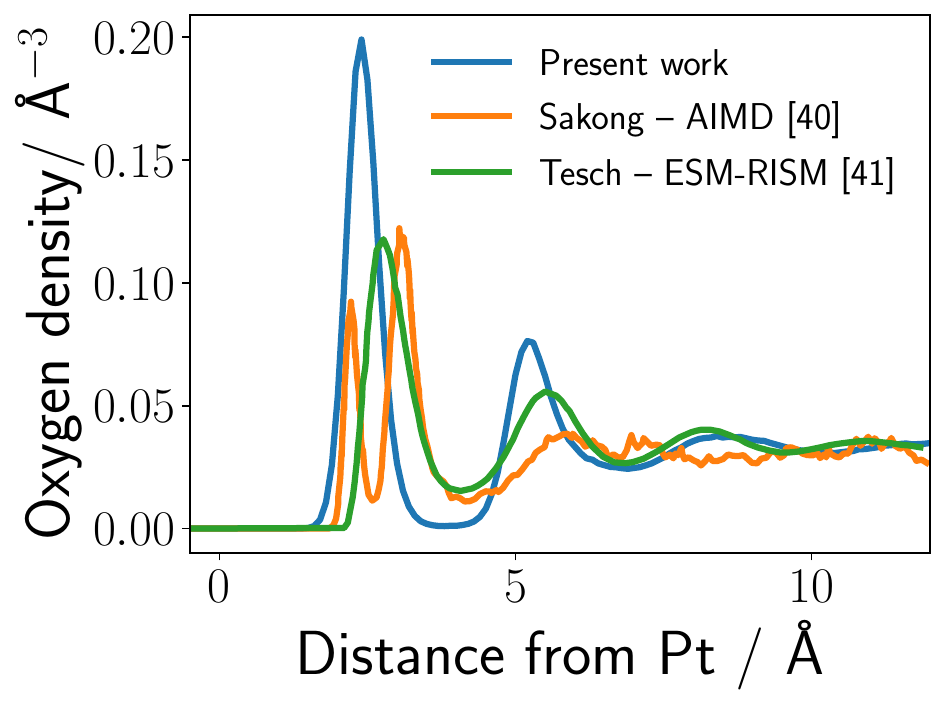}
                \caption{\label{fig:zdens_ptwat_o}Density distributions of water oxygen from different levels of theory.\cite{sakong,tesch}}
            \end{figure}

            We asses the used FF in terms of its ability to adequately represent the structure of water in contact with the platinum surface. This interface has been investigated in multiple studies with various computational approaches, including \textit{ab initio} MD (AIMD) which offers accurate atomic scale insight, with which we can compare our results.\cite{sakong, tesch} We find that the distribution of oxygen atoms of water from our simulations is in good agreement with literature results, as shown in Fig.\ \ref{fig:zdens_ptwat_o}. The AIMD simulations of Sakong and Groß yielded a water configuration with a double peak feature between 2--4\,\AA\ away from the surface.\cite{sakong} Our results, as well as the results of Tesch et al., obtained using the effective screening medium - reference interaction site method (ESM-RISM) without treating water molecules in the first adsorption layer explicitly, do not show this feature.\cite{tesch} Comparing the position of the peak closest to the Pt surface, we note that the one obtained in our studies is slightly closer to the surface compared to the other studies. The utilized force field slightly overestimates the adsorption of water onto the Pt surface, which we will take this into consideration for the main results of the presented work.
                    
\begin{figure}[!ht]
                \centering
                \includegraphics[width=.4\textwidth]{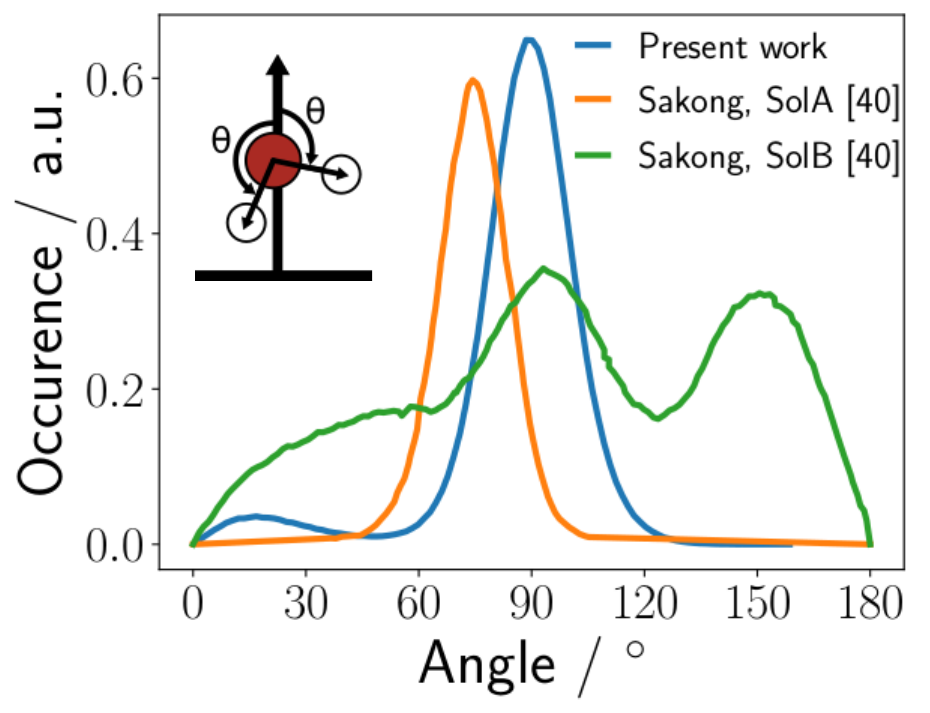}
                \caption{\label{fig:ang_wat_ptwat}Distributions of water OH bond angles to surface normal from this study (blue) and AIMD data (orange, green) taken from Ref.\cite{sakong}.}
            \end{figure}

            Sakong and Groß also analysed the orientation of the water molecules that are less than 4\,\AA\ away from the platinum surface. They analysed the populations corresponding to the two peaks in Fig.\ \ref{fig:zdens_ptwat_o} separately. The population belonging to the closest peak is denoted as SolA, the one belonging to the second peak as SolB. Adopting their analysis method of calculating the angle between the OH bonds with the surface normal, we compare the orientation of the molecules with a distance of less than 4\,\AA\ from the surface found in our simulations to their distributions in Fig.\ \ref{fig:ang_wat_ptwat}. In our simulations, the water molecules adsorb in a flat configuration as indicated by the main peak around 90\,$^\circ$. The small second peak that is located at around 17\,$^\circ$ indicates a few hydrogen atoms are pointing towards the second water layer. The second peak of the SolB data around 150\,$^\circ$, corresponding to hydrogen atoms that point towards the platinum surface, is not observed in our data. With this analysis we note that care has to be taken in interpreting structural information for the direct adsorption layer atop the platinum surface obtained with the used FF.

        \subsection{Structural analysis of the Pt--water--Nafion interfaces}

            \begin{figure*}[!ht]
                \centering
                \includegraphics[width=\textwidth]{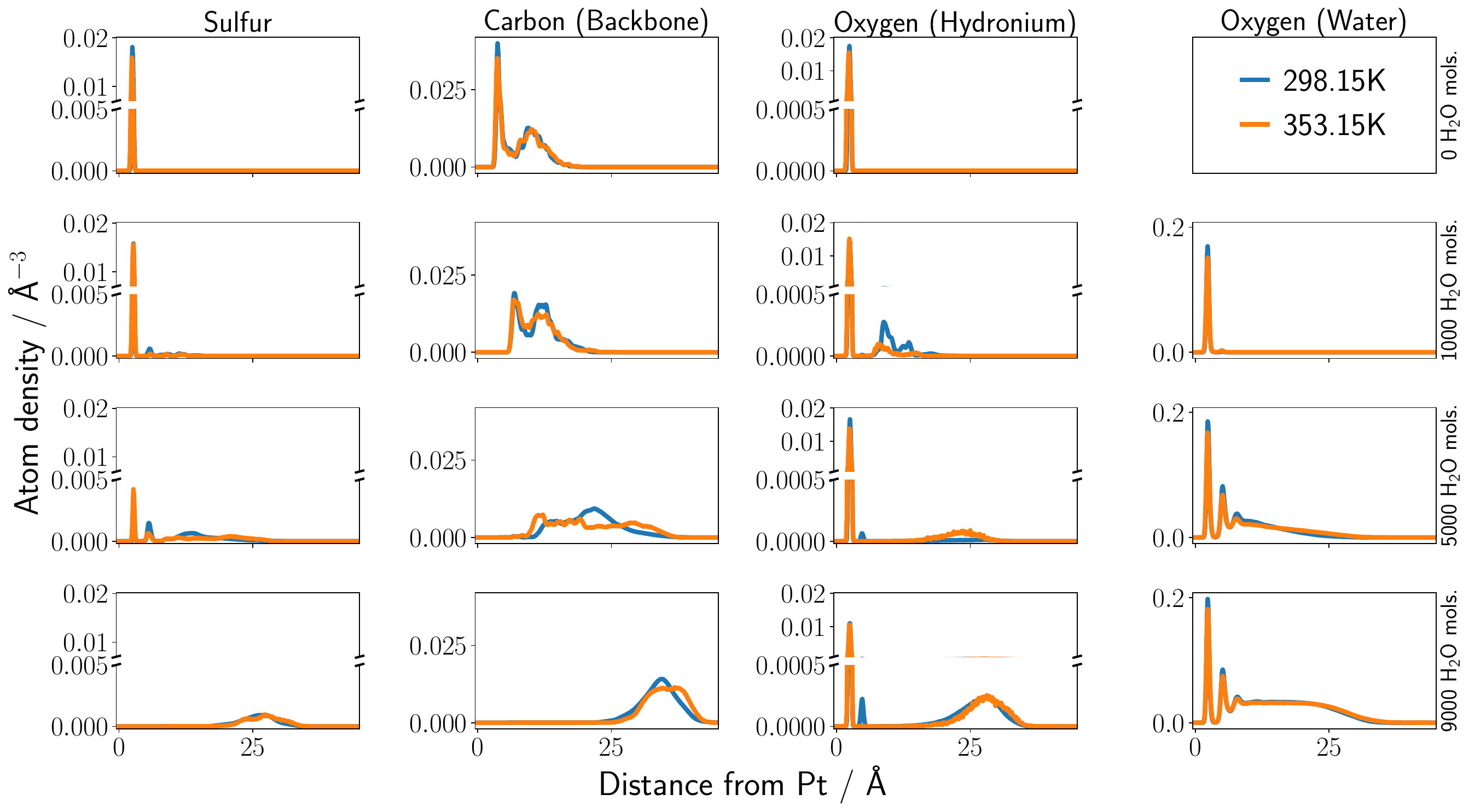}
                \caption{\label{fig:zdens_naf_fl_woc}Densities along Z direction from the Pt surface for selected atom species (column wise) of the `dense' layer with different amounts of added water molecules (row wise). Note the broken axes in columns 1 and 3.}
            \end{figure*}
            
            As the first analysis step of the platinum-water-Nafion systems we turn towards the structural features of the systems. Figure~\ref{fig:zdens_naf_fl_woc} shows the Z-density profiles of sulfur and backbone carbons of Nafion, as well as the profiles for the oxygen atoms of the hydronium and water molecules. For the sulfur atoms, we observe a peak close to the surface for all simulations except those with 9000 water molecules added, showing that there is contact between the platinum surface and the Nafion sulfo groups for the systems with a small number of water molecules. A second peak is observed for simulations with 5000 added water molecules at both temperatures. In the case of the simulation with 5000 added water molecules at RT the peak height of the second peak is even higher than the first. From the snapshot of this system in Fig.\ \ref{fig:snaps} we can see that the two peaks are caused by modulations that render some strands of the Nafion layer closer to the platinum surface than others. The waviness of the ionomer thin film in the X-Y plane is a result of the intricate interactions at play at this interface. The water is bound on one side by the platinum and on the other side by the Nafion strands, whose backbone is highly hydrophobic. Therefore, the area between water and the Nafion backbone is minimized during the simulation.

\begin{figure*}[!ht]
        \centering
        \includegraphics[width=\textwidth]{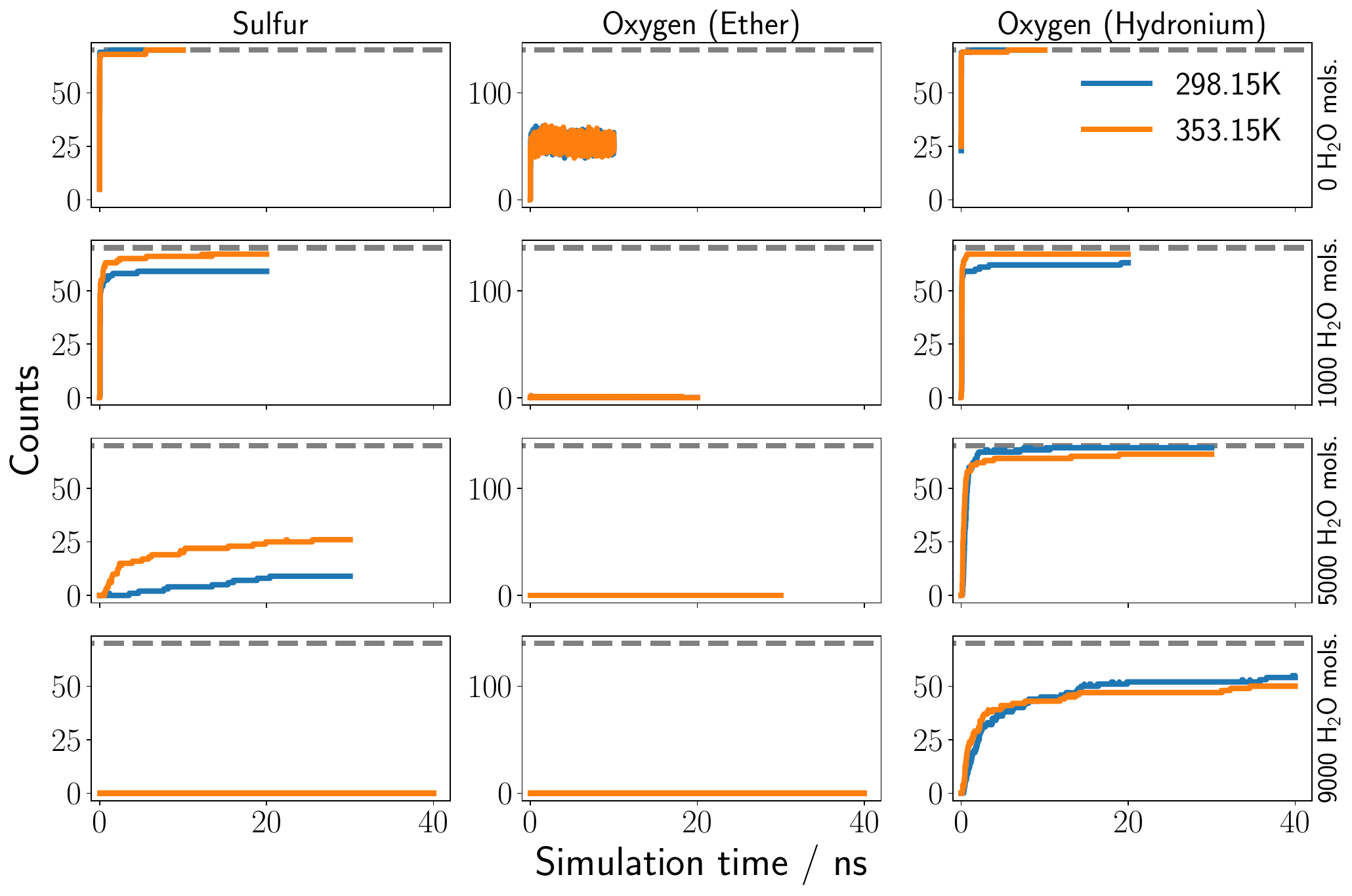}
        \caption{\label{fig:contacts}Temporal evolution of the number of atoms that are within 4\,\AA\ of the platinum surface. Total number of each species shown as dashed grey line.}
    \end{figure*}

            The density profiles of backbone carbon atoms show the expected behaviour of the Nafion layer being more separated from the platinum surface with increased water content. In simulations with 5000 added water molecules, the distributions are broader compared to the other simulations, which is a direct result of the wavy structure described above.
            For the density profiles of the hydronium oxygen atoms we notice that in all cases the first large peak is located in close proximity to the platinum surface, indicating a strong attraction of the hydronium ions to the surface. The temporal evolution shows that the number of ions at the platinum surface steadily increases with time in all simulations (SI\ Fig.\ \ref{fig:contacts}). This strong interaction of the hydronium ions with the platinum surface was ascribed to polarization of the metal surface due to the charged hydronium ions by Brunello et al.\ when they fitted the FF.\cite{brunello} The presence of hydronium ions close to the platinum (111) surface is also supported by AIMD simulations, where spontaneous formation of hydronium in the water film less than 5\,\AA\ away from the platinum surface has been observed.\cite{sakong_b} In their work, Sakong and Groß observed hydronium ions with a population probability of 10\% per time step.\cite{sakong_b} For a system with an implicit HCl concentration of 0.1\,mol/L, Tesch et al. also observed hydronium ions near the platinum surface at the potential of zero charge (PZC).\cite{tesch} In their work, the first adsorption layer had increased concentration of $\sim$0.113\,mol/L. With an increase in the applied potential, the density of hydronium ions near the surface diminishes. As simulations discussed so far use an uncharged platinum slab they can be regarded as being performed at the PZC. Thus, the accumulation of hydronium ions is appropriate, however, the fact that $\sim$70\% of the hydronium ions are within 4\,\AA\ from the surface is not justified by DFT data in literature and can be regarded as overfitting of the utilized FF.
            The density plots of the water oxygen atoms show only a single peak when 1000 water molecules are added, relating to only a singular water layer at the surface. The peak height is comparable to that of the first peak of the plots with higher water content. In the case of 5000 added water molecules, we see two additional peaks as well as a tail that extends to $\sim$27\,\AA. Adding further water molecules extends this tail end to $\sim$35\,\AA\ from the surface. While these values show the maximum distance to the surface where water is present they are not a direct value of the water film thickness, as the Nafion layer is extending into the water film as well.
            
\begin{figure}[!ht]
                \centering
                \includegraphics[width=.49\textwidth]{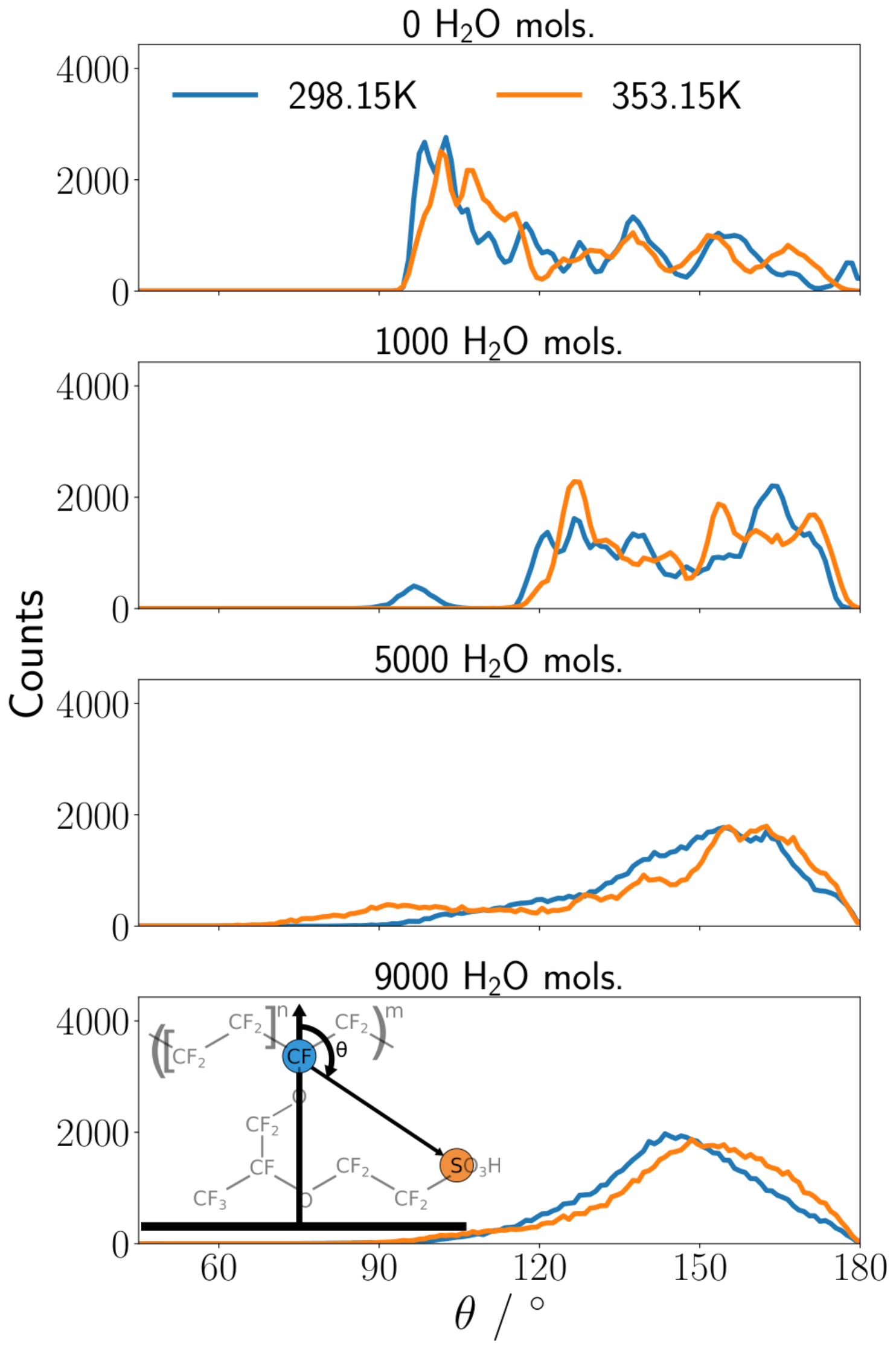}
                \caption{\label{fig:sc_ang_fl}Side-chain angles histograms of the `dense' layer systems with different water content.}
            \end{figure}

\begin{figure*}[!ht]
        \centering
        \includegraphics[width=.8\textwidth]{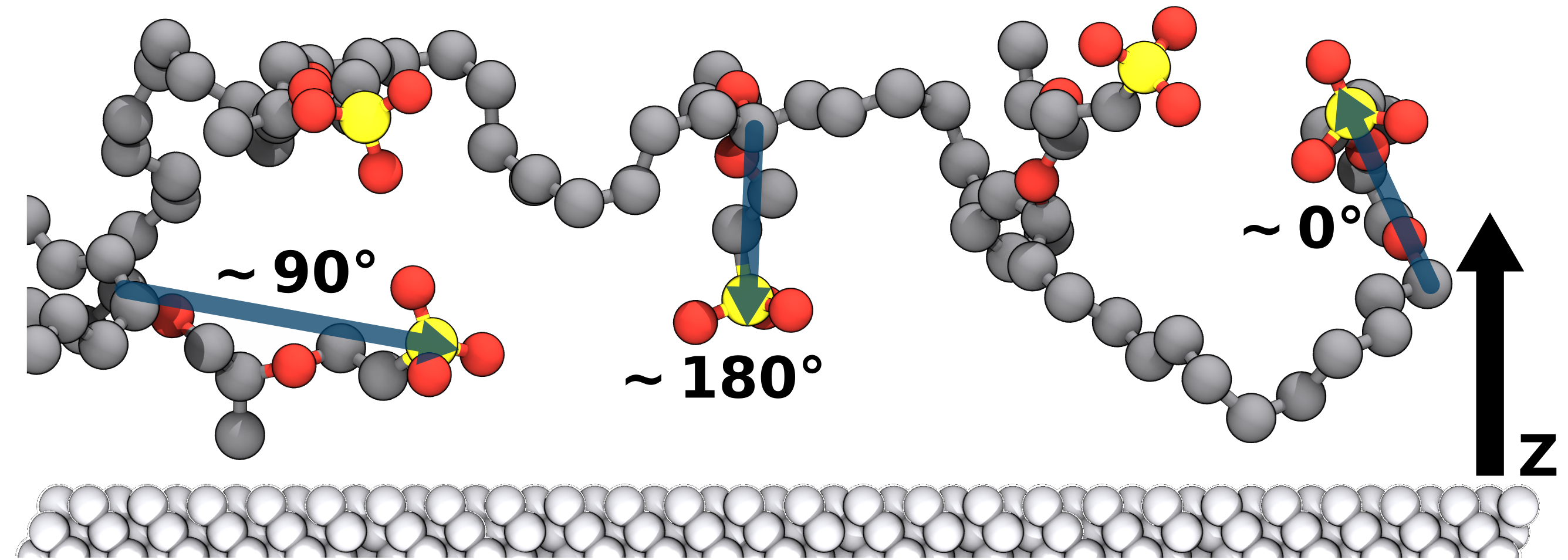}
        \caption{\label{fig:sc_ang_expl}Visual example of side-chain angles w.r.t. the platinum surface.}
    \end{figure*}

            A key structural feature of Nafion at the catalyst interface is the orientation of the side-chains with respect to the surface. Here, we describe the orientation via the angle between the vector from the carbon atom that connects to the side-chain to the sulfur atom of the side-chain, and the catalyst-surface normal (cf.\ SI\ Fig.\ \ref{fig:sc_ang_expl}). The resulting angle distributions are shown in Fig.\ \ref{fig:sc_ang_fl}. We observe that the side-chains tend towards a downward orientation to the surface with increased water content. Without added water, the distributions are spread over a broad range from 100\,${^\circ}$ to 180\,${^\circ}$. As more water is added between the platinum surface and the Nafion thin film, this range narrows slightly to 120--180\,${^\circ}$. For the systems with 5000 and 9000 added water molecules the maxima of the angular distributions are at $\sim$160\,${^\circ}$ and $\sim$150\,${^\circ}$, respectively, showing a preference of the side-chains to be orientated towards the platinum surface.

            Another key observation is that the shape of the distribution curves becomes smoother with an increase in water content. As the Nafion side chains are in direct contact with the surface in the low water content systems (cf. Fig.\ \ref{fig:zdens_naf_fl_woc}) they are less mobile, and the range of angles a specific side chain can explore during the simulation is constrained. 

        \subsection{Estimating equilibrium water layer thickness}
            Utilising the energies of the different systems obtained during simulations, the equilibrium water film thickness can be estimated. The favourable water content can be estimated by relating the total energies $E$ of two systems $A$ and $B$ with water contents $W_A$ and $W_B$, where $W_B$ is larger than $W_A$ via
            \begin{align}
                \label{eq:delta_e}
                \Delta E = \frac{E_B - \left(E_A + ((W_B-W_A) \cdot E_\mathrm{{H_2O}} )\right)}{W_B-W_A},
            \end{align}
            where $E_{H_2O}$ is the energy of a water molecule obtained from simulations of bulk water.
            With the normalization per water molecule it can be considered an approximation for the chemical potential of water, where instead of the insertion of one molecule the average of 1000 or 4000 molecules is considered.
            The energy differences, as listed in Tab.\ \ref{tab:energies}, show a minimum for both temperatures when 1000 water molecules are added into the dry system. This minimum energy is negative indicating, that the solvation of the system is energetically favoured. If more water is added, the energies become positive indicating instablility of the water film. In the initial case of water being introduced into the system the energy is negative due to the strong interaction of the water molecules with the platinum surface as well as the electrostatic contributions from the interactions of water with the hydronium and sulfonic groups. Once these interactions are established adding more water is energetically unfavourable since the nanoconfinement constrains the water arrangement and thus the hydrogen bond network. 
            From the explored water ranges we can draw a rough preliminary conclusion that a stable water film of less then 13\,\AA, comparable in thickness to the experimental values of Chabot et al.\cite{chabot_b}, forms in the gap between platinum and Nafion. By performing a more thorough screening, more precise values of the water film thickness will be obtained. Using grand canonical methods will allow the inclusion of entropic effects so far not considered.\cite{kanduc_m}

            \begin{table}
                \centering
                \caption{\label{tab:energies}Difference in energy per water molecule of the systems w.r.t. to the system with the next lower water content, calculated with Eq.\ \ref{eq:delta_e}.}
                \begin{tabular}{ccc}
                    \toprule
                    Water diff.& \multicolumn{2}{c}{$\Delta$E [eV H\textsubscript{2}O\textsuperscript{-1}]}\\
                    & 298.15\,K& 353.15\,K\\ \midrule
                    1000--0&    -0.634$\pm$0.010&  -0.717$\pm$0.014\\
                    5000--1000&  0.0225$\pm$0.003&  0.0285$\pm$0.004\\
                    9000--5000&  0.0375$\pm$0.004&  0.0338$\pm$0.006\\ \bottomrule
                \end{tabular}
            \end{table}

        \subsection{Electrostatics}

            After evaluating the structural features and water layer thickness of the uncharged system, we also investigate the effect of the water content on the electric surface potential of the platinum slab. The potential is calculated as described in the methods section for both uncharged systems and systems with a charged Pt surface.

            \begin{figure*}[!ht]
                \centering
                \includegraphics[width=.75\textwidth]{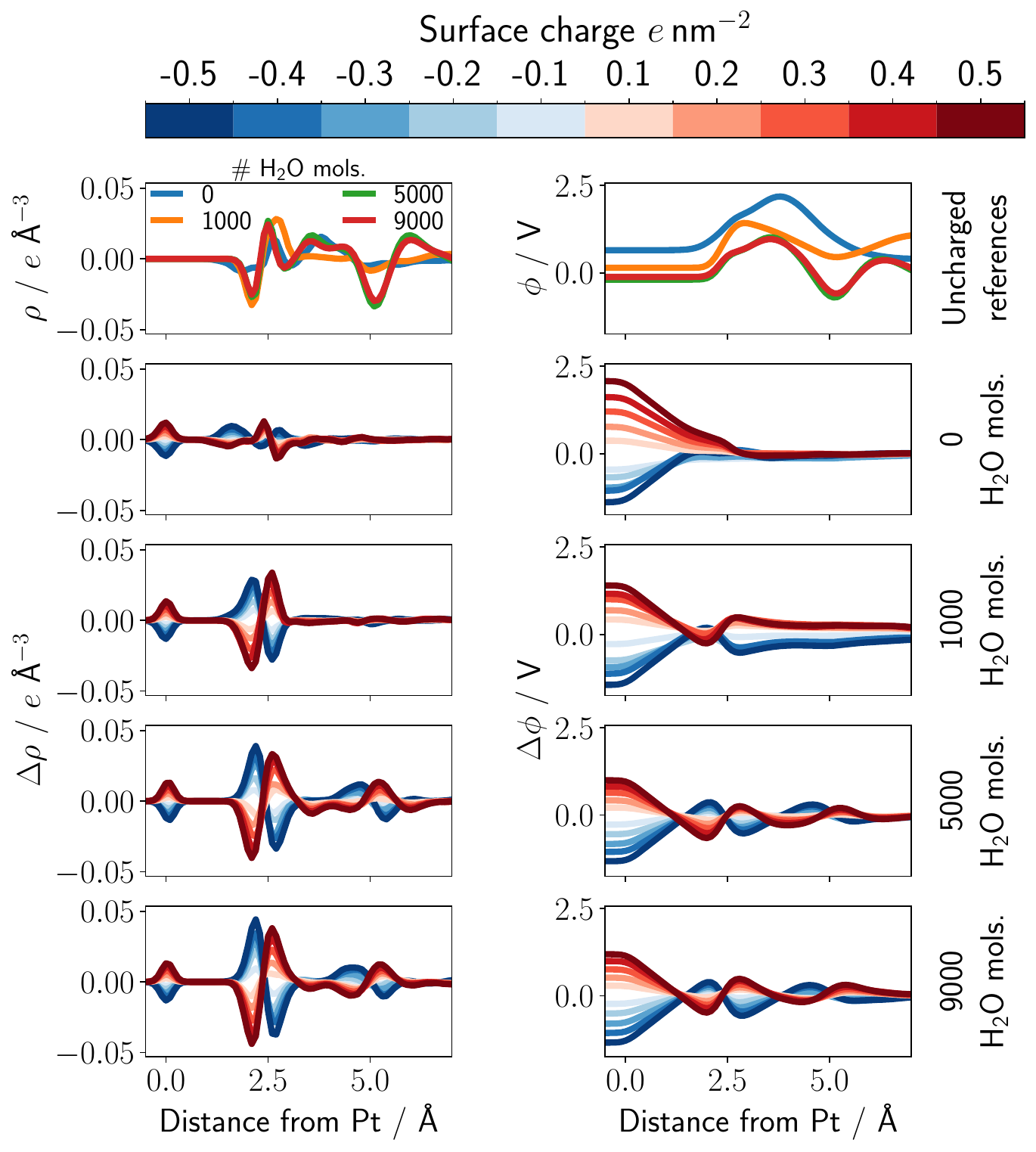}
                \caption{\label{fig:es_cdens_pot}Comparison of charge densities (left) and electrostatic potentials (right) between uncharged and charged systems. Top row are references from the uncharged simulations for each water content. Rows 2--5 are differences in charge density ($\Delta\rho$ (left)) and potentials ($\Delta\phi$ (right)) between the charged systems and the uncharged reference for 0, 1000, 5000 and 9000 added water molecules, respectively.}
            \end{figure*}

\begin{figure*}[!ht]
        \centering
        \includegraphics[width=\textwidth]{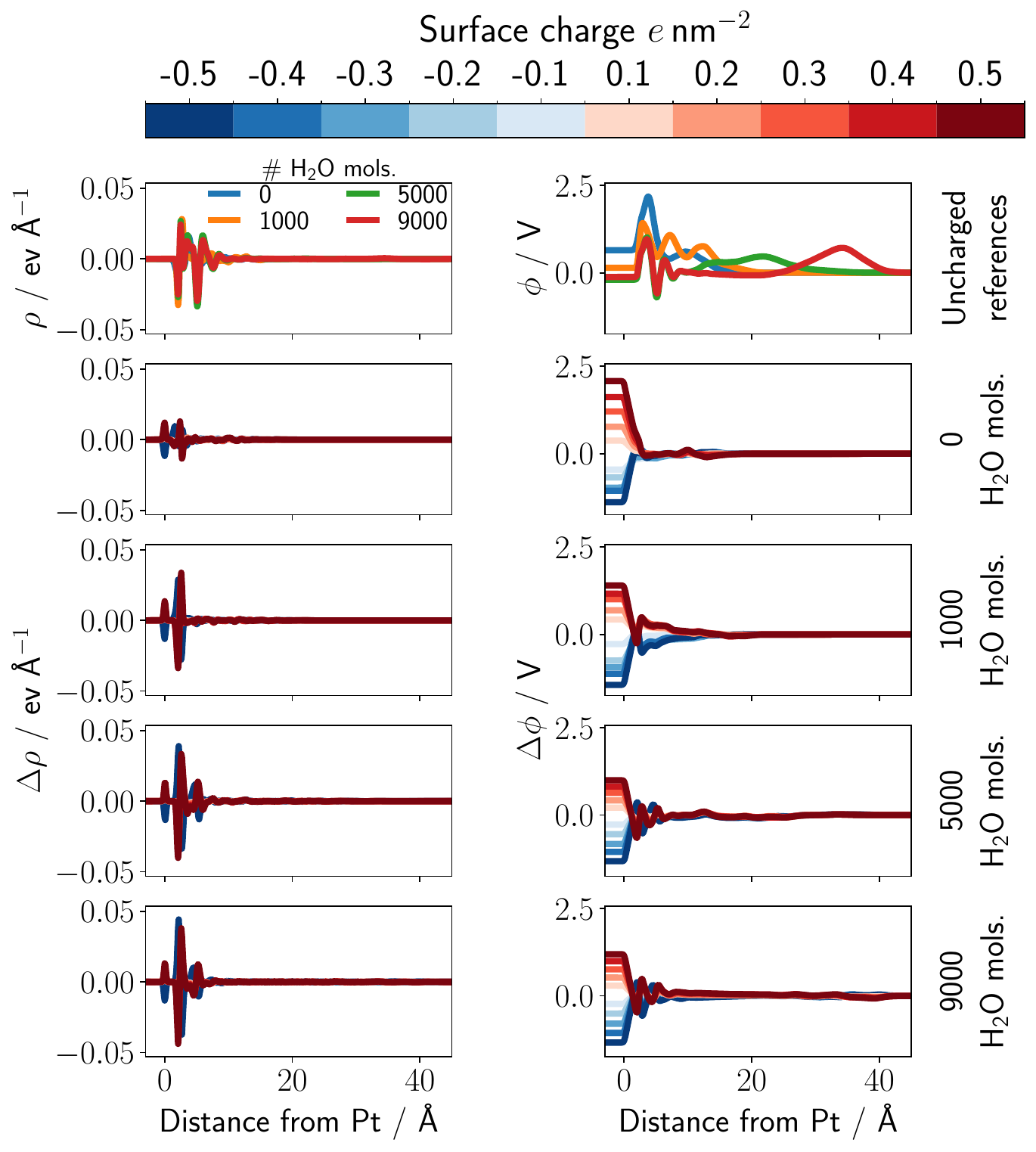}
        \caption{\label{fig:es_cdens_pot_long}Comparison of charge densities (left) and potentials (right) between uncharged and charged systems. Top row are references from the uncharged simulations for each water content. Rows 2--5 are differences in charge density ($\Delta\rho$ (left)) and potentials ($\Delta\phi$ (right)) between the charged systems and the uncharged reference for 0, 1000, 5000 and 9000 added water molecules, respectively.}
    \end{figure*}

            The difference in charge densities and potentials of the charged simulations with respect to the uncharged ones are shown in Fig.\ \ref{fig:es_cdens_pot}, with:
            \begin{align}
                \begin{split}
                    \Delta\rho(z) &= \rho^{chg}(z)-\rho^0(z) \\
                    \Delta\phi(z) &= \phi^{chg}(z)-\phi^0(z),
                \end{split}
            \end{align}
            where $\rho^{chg}(z)$ and $\phi^{chg}(z)$ are the charge density and electrostatic potential of a charged system at position $z$, respectively, and the superscript $0$ denotes the uncharged equivalents.
            The plots only show the region near surface as there is no difference in potential with a change in surface charge beyond that region for all systems (cf.\ SI\ Fig.\ \ref{fig:es_cdens_pot_long}). The first peak in all systems, located at 0\,\AA\ distance, corresponds to the change in charge of the surface platinum atoms. The charge densities for systems with 5000 and 9000 added water molecules are overall similar with slight differences in peak heights. In the case of 1000 added water molecules, differences compared to the uncharged case are only visible within a distance of less than 4\,\AA\ from the surface. An inflection point showing a shift in atomic positions is present in the charge density plots for all systems at $\sim$2.5\,\AA. These inflection points indicate a shift of the adsorbed species either towards or away from the surface. For simulations with 5000 and 9000 added water molecules, an additional inflection point occurs at $\sim$5\,\AA\ that represents the second adsorption layer.

\begin{figure}[!ht]
                \centering
                \includegraphics[width=0.5\textwidth]{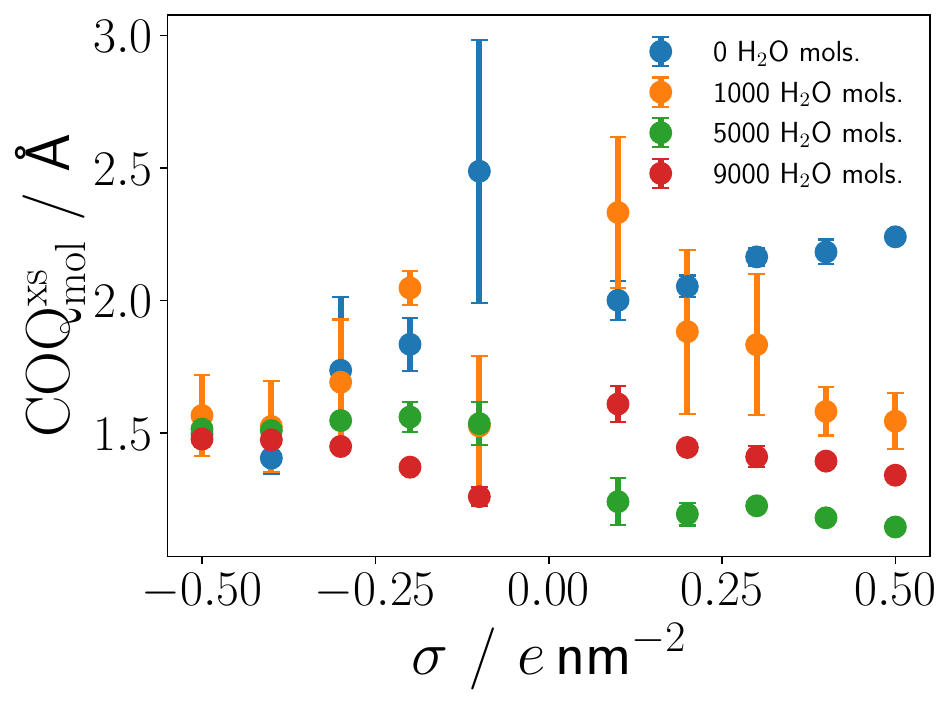}
                \caption{\label{fig:coq}Centers of molecular excess charge (cf. Eq.\ \ref{eq:coq}) per water content with respect to surface charge.}
            \end{figure}

            To gain a clear picture of the effect of the added surface charge on the properties of electrolyte, we utilse the center of molecular excess charge ($\mathrm{COQ_{\mathrm{mol}}^{\mathrm{xs}}}$). The basis for this is the center of charge that is a well-established concept in the literature, with the generalisation to the ionic charges similar to the capacitive compactness.\cite{luque, guerro_garcia} To calculate the $\mathrm{COQ_{\mathrm{mol}}^{\mathrm{xs}}}$ the following equation is used
            \begin{equation}
                \label{eq:coq}
                \mathrm{COQ_{\mathrm{mol}}^{\mathrm{xs}}} = \frac{\displaystyle\int \left(\rho^{chg}_{mol}(z)-\rho^0_{mol}(z)\right)\cdot z \,dz}{\displaystyle\int \left(\rho^{chg}_{mol}(z)-\rho^0_{mol}(z)\right) \,dz},
            \end{equation}
            where $z$ is the distance from the platinum surface, $\rho^{chg}_{mol}(z)$ and $\rho^0_{mol}(z)$ are the molecular charge densities, i.e., without considering the platinum atoms, averaged over the X-Y-plane at position $z$ of the charged and uncharged systems, respectively. By plotting the resulting center positions against the surface charge, the shift of the molecular excess charge can be tracked. Fig.\ \ref{fig:coq} shows that the $\mathrm{COQ_{\mathrm{mol}}^{\mathrm{xs}}}$ converges to a similar value for all systems as  along the branch of negative surface charge as the absolute charge increases. On the branches for the positive surface charges this convergence is not observed. Noteabley, the sign of the surface charge at which the center is closest to the platinum surface is different depending on the water content. In the case of 5000 and 9000 added water molecules the $\mathrm{COQ_{\mathrm{mol}}^{\mathrm{xs}}}$ is closest to the surface for positive charges. For the system with 1000 added water molecules, the positions are equal for positive and negative surface charges within the respective errors. For zero added water the clostest postion is observed for negative charges. 

            The electrostatic potential plots shown in the second column in Fig.\ \ref{fig:es_cdens_pot} for the systems with 5000 and 9000 added water molecules are very similar, with only minute differences in peak heights. One structural difference between these two systems is the presence of sulfonic acid head groups that are adsorbed to the surface when 5000 water molecules were added. The different peak heights can most likely be attributed to the change in electrostatic interaction between these sulfonic acid head groups and the charged surface. 
            
\begin{figure}[!ht]
                \centering
                \includegraphics[width=.49\textwidth]{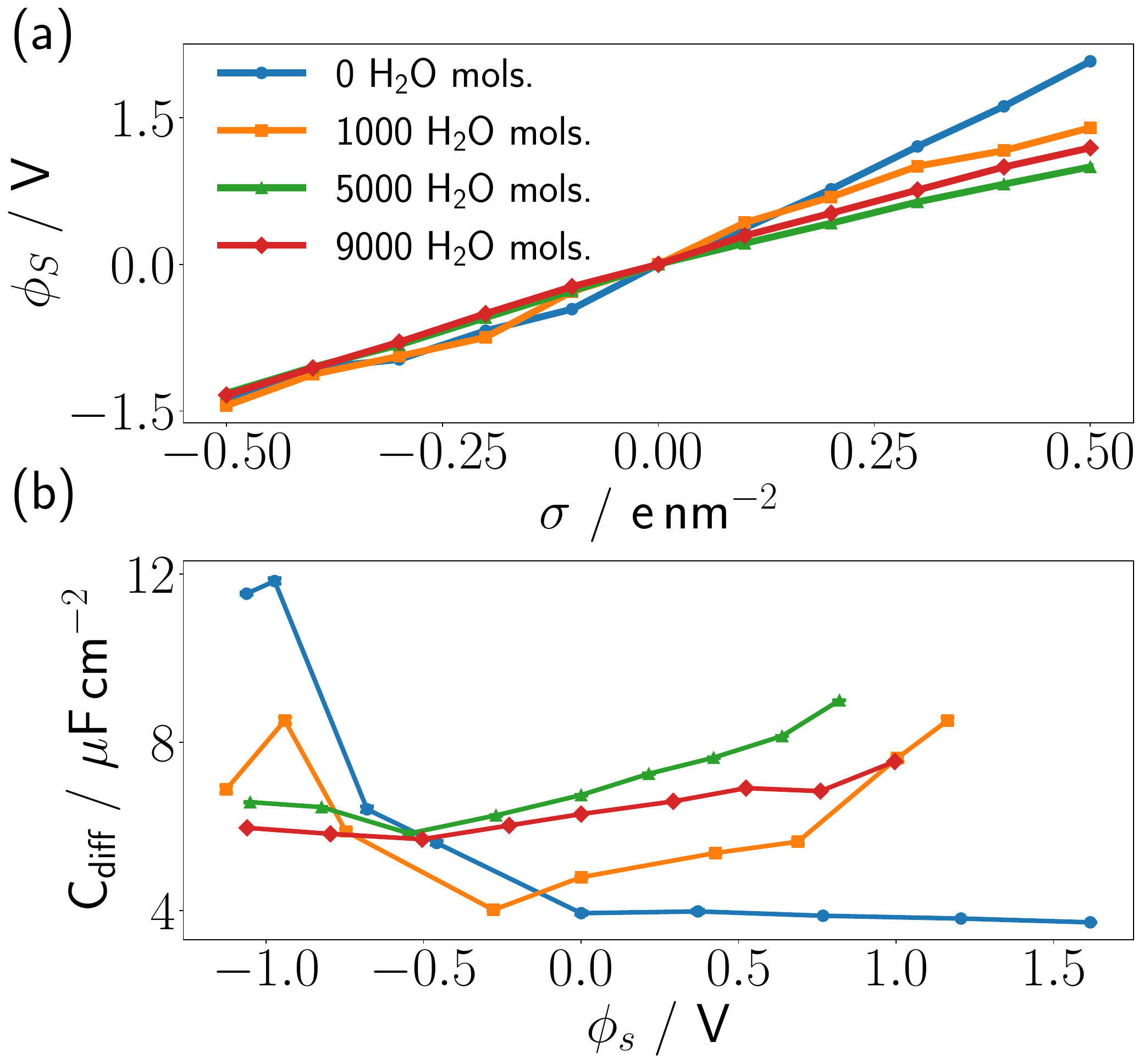}
                \caption{\label{fig:deltapot_per_water}(a) Electric surface potentials $\phi_s$ as a function of surface charge for each water content. The potential reference is chosen such that the uncharged systems have $\phi_s$=0. (b) Differential capacitances C\textsubscript{diff} as a function of electric surface potential for each water content. Calculated using Eq.\ \ref{eq:dlcap} using neighbouring charge steps and $\phi_s$ values as described in Sec.\ \ref{sec:esp_dl}.}
            \end{figure}

            In the systems where no water has been added, the potential plots of the simulations with negatively charged surfaces are overlapping to a higher extent than those of positively charged surface systems. As a result, the spreads of electric surface potentials for negative and positive surface charges deviate greatly. This kind of deviation in magnitude of $\phi_s$ between the opposing surface charges is also observed in all other systems, however, in these simulations the larger magnitude is observed when the surface is negatively charged. When plotting $\phi_s$ against the surface charge, as shown in Fig.\ \ref{fig:deltapot_per_water}(a), we thus observe asymmetry in the plots for negative and positive values of $\sigma$. It is reminded that, for each system, the zero potential reference was chosen to correspond to the respective uncharged surface state (i.e., the PZC which is $\sim$0.23\,V vs. SHE).\cite{xu_p} The relation of the plots to one another closely matches the relation observed in Fig.\ \ref{fig:coq}. 

            In Fig.\ \ref{fig:deltapot_per_water}(b) the differential capacitances, calculated with Eq.\ \ref{eq:dlcap}, are shown. 
            The respective values are in the range of 4--12\,$\mu$F\,cm$^{-2}$, which is comparable to the values obtained when using the data of Zhang et al., although their simulations were performed using a $\sim$2.5\,nm thick hydrated Nafion film.\cite{zhang} The plot shows asymmetric behaviour that changes with increased water content. Starting from the simulations with no water, we observe a maximum on the negative potential side and a plateau on the positive side. Once water is introduced, the negative side of the capacitance plot tends towards a plateau and the positive side shows a positive slope.
            
\begin{figure*}[!ht]
                \centering
                \includegraphics[width=.75\textwidth]{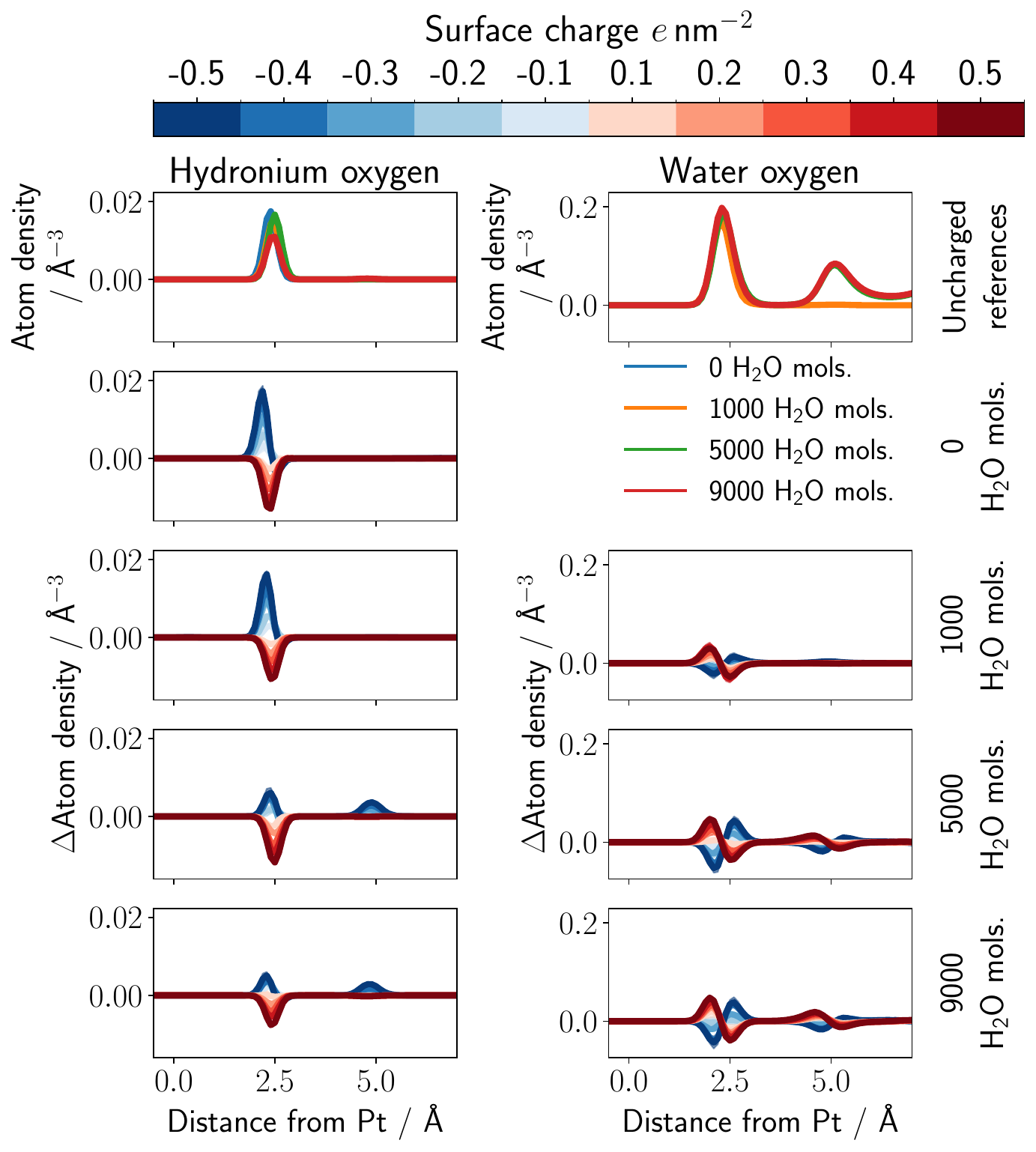}
                \caption{\label{fig:es_hyd_wat}Comparison of Z densities between uncharged and charged systems of hydronium oxygen (left) and water oxygen (right). Top row are references from the uncharged simulations for each water content. Rows 2--5 are differences in the occurrences of the specific atom species between the charged systems and the uncharged reference for 0, 1000, 5000 and 9000 added water molecules, respectively.}
            \end{figure*}

            As the differential capacitance shows how the electrostatic potential changes due to changes in the applied surface charge, and the electrostatic potential is derived from the charge denisty, the Fig.\ \ref{fig:deltapot_per_water}(b) gives an indication for the restructuring of the interfacial region upon adding water and varying the electrode potential. To elucidate this restructuring, we show the change in Z densities of the oxygen atoms of the hydronium and water molecules in Fig.\ \ref{fig:es_hyd_wat}. As already discussed for the uncharged reference, the hydronium ions are strongly adsorbed to the surface yielding a large peak near the surface that accounts for $\sim$70\% of the hydronium ions. Once the surface charge of the platinum slab is altered, the peak either de- or increases depending on the sign of the surface charge when hydronium ions are deleted or added, respectively. For the systems with 0 or 1000 added water molecules, the plots are very similar; however, as more water is added to the system, a second peak appears at $\sim$5\,\AA\, but only if the surface is negatively charged, i.e., for added hydronium ions. The stable second layer is formed once 30 or more hydronium ions are added, corresponding to surface charges of $-$0.3\,$e$\,nm$^{-2}$ or below. Thus, the presence of the second adsorption layer can be ascribed to crowding effects in the first adsorption layer of the hydronium molecules. This may stem, for example, from the grwoth of electrostatic attraction of both the water molecules and hydronium ions in the first adsorption layer with the surface, which in turn prevents uptake of further hydronium ions into the first adsorption layer. With the second adsorption layer, the electric double layer thickness increases, causing the differential capacitance to decrease in the negative potential region.
            For the positive potentials the observation is opposite. Here, the differential capacitances increase when water is added to the system. The Nafion layer more closely approaches the platinum surface as the water content decreases. Sulfonate anions at the Nafion side-chains will accumulate near the Pt surface at lower water content and in the region of positive surface charge density, leading to a reduction of the differential capacitance.

\begin{figure}[!ht]
        \centering
        \includegraphics[width=\textwidth]{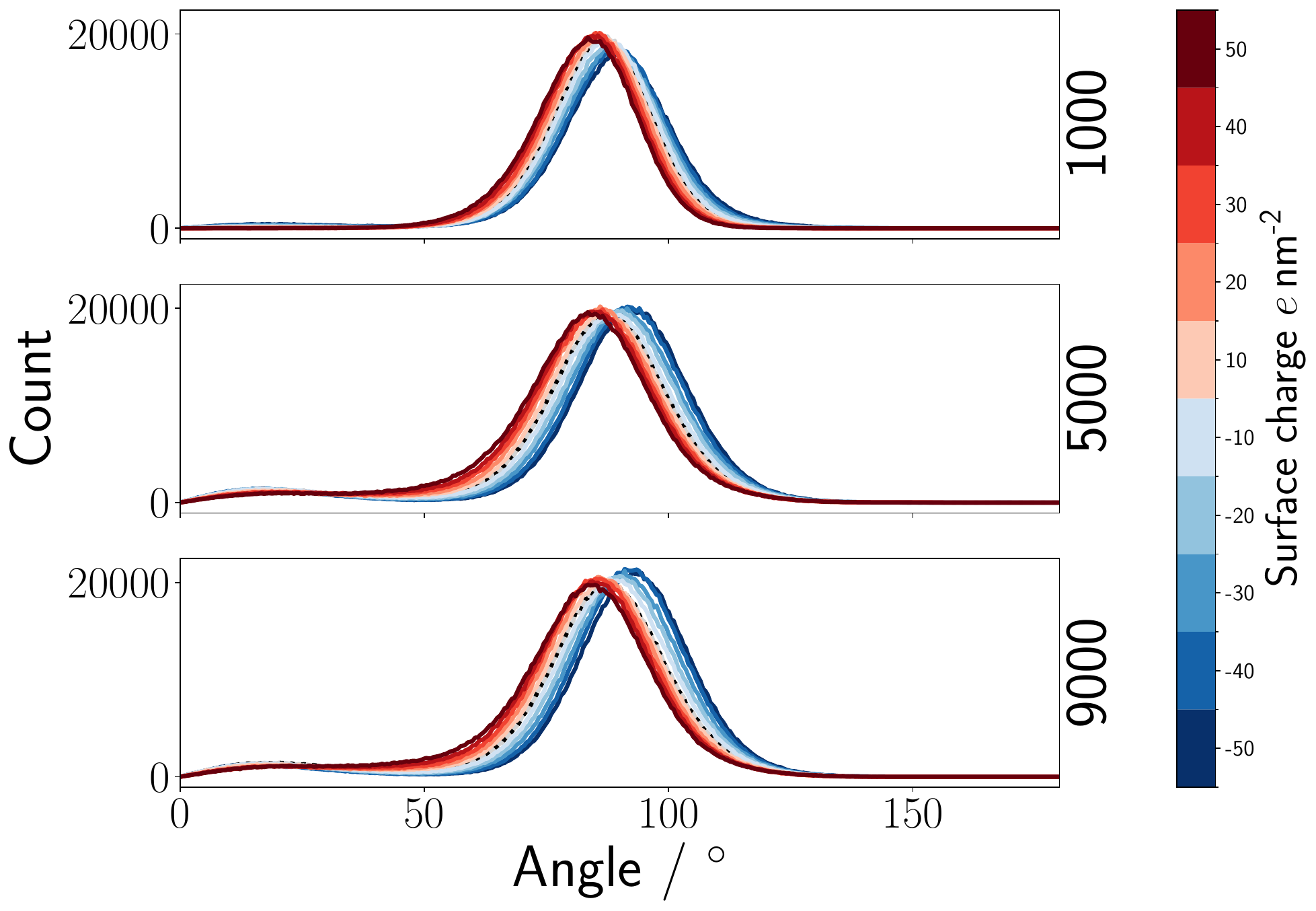}
        \caption{\label{fig:ang_si}Distribution of water dipole angles w.r.t. the surface normal for the charged systems.}
    \end{figure}

            The changes in the Z-density profiles of the water oxygen atoms, as shown in Fig.~\ref{fig:es_hyd_wat} (right), are monotonic to one another with no differing features as seen for the hydronium Z densities. The shifts of the water molecules, as indicated by the inflection point and the change in water orientation (cf. SI\ Fig.\ \ref{fig:ang_si}), may play a small role in the differential capacitance, but a large influence can be ruled out. The main influence is the amount of water present, which leads to the shift in magnitude of the diferential capacitances.
            
            The electrostatic potential, and thus the differential capacitance, of the platinum-nafion interface region is ultimately largely influenced by the ammount and spatial distribution of hydronium ions. Due to the strong accumulation of hydronium ions near the platinum surface the surface charge of the platinum is screened from the Nafion anion groups. With an added water film separating the Nafion from the surface the differential capacitance curves are smoothed out as the restructuring is limited. These minimal changes are trackable by considering the $\mathrm{COQ_{\mathrm{mol}}^{\mathrm{xs}}}$.
            With the large influence of hydronium adsorption, follow-up work is needed to scrutinze the adsorption behaviour.

  \section{Conclusion}
    We characterized the structural and electrostatic features of a catalyst--ionomer model interface constructed with a platinum (111) surface, varying degrees of water content and a Nafion thin film. A dense and relatively uniform Nafion thin film was generated by employing Voronoi tesselation.
    From the thermodynamic stability evaluation, we infer the water film thickness between platinum and Nafion to be less than 13\,\AA. Follow up simulatios are planned to investigate this further by employing grand canonical Monte Carlo simulations.
    In our electrostatic analysis we utilised the center of molecular excess charge $\mathrm{COQ_{\mathrm{mol}}^{\mathrm{xs}}}$ as a means of tracking changes in the electrlyte due to added surface charge. These centers show correlation with the electrostatic potential, while providing additional insight into the structure of LRE. By computing the differential capacitances for systems with charged Pt surfaces, we observe peculiar trends in the charging behaviour around the PZC. We could identify changes in the accumulation of hydronium ions as the source of this behaviour. Another effect shaping the capacitance response will be the close approach of sulfonate anions to the platinum surface for systems with small water content and positive platinum surface charge.

    Even in the uncharged case a large portion of hydronium ions was adsorbed to the surface, and by adding further ions into the system to balance the applied negative charge to the platinum surface, a second adsorption layer is formed. 
    The presented workflow will be the basis for future studies employing methods with increased accuracy, or for exploring other electrochemically relevant metal-ionomer interfaces with PFAS-free ionomers. 
    Follow-up work will need to address the effect of adsorbed hydrogen, hydroxide, and oxygen on the platinum surface, as is known from experiment\cite{gomez,rizo}, as well as reevaluate the force field in terms of adsorption strength of the hydronium ions.

  \section*{CRediT authorship contribution statement}
    \textbf{Dustin Vivod}: Writing --- original draft, Visualization, Validation, Methodology, Investigation, Formal Analysis, Conceptualization. \textbf{Binny A. Davis}: Conceptualization, Methodology, Formal analysis. \textbf{Tobias Binninger}: Writing --- review \& editing, Supervision, Resources, Project Administration. \textbf{Michael Eikerling}: Writing --- review \& editing, Supervision, Resources, Project Administration, Funding acquisition.

  \section*{Declaration of competing interest}
    The authors declare no competing financial interest.

  \section*{Acknowledgments}
    The authors gratefully acknowledge the Gauss Centre for Supercomputing e.V. (\url{www.gauss-centre.eu}) for funding this project by providing computing time through the John von Neumann Institute for Computing (NIC) on the GCS Supercomputer JUWELS at J\"ulich Supercomputing Centre (JSC). The authors acknowledge the financial support from the Federal Ministry of Science and Education (BMBF) under the German-Canadian Materials Acceleration Centre (GC-MAC) grant number 01DM21001A, and HIGHLANDER which receives funding from the Clean Hydrogen Partnership under grant agreement number 101101346.

    \printbibliography

  \pagebreak
  \section*{Supporting Information}

    \begin{table*}
        \caption{\label{tab:voronoi_si}Minimum, maximum, and median areas calculated from the Voronoi tesselation for all simulation systems.}
        \begin{tabular}{lccccc}
            \toprule
            \# Water& Nafion& Temperature& Minimum& Maximum& Median\\
            molecules&layer&[K]&area&area&area\\
            &type&&[nm\textsuperscript{-2}]&[nm\textsuperscript{-2}]&[nm\textsuperscript{-2}]\\ \midrule
            \multirow[t]{4}{*}{0}& \multirow[t]{2}{*}{square} & 298.15K & 1.27 $\pm$ 0.380& 54.3 $\pm$ 5.47& 12.0 $\pm$ 0.086\\
            &  & 353.15K & 1.02 $\pm$ 0.556& 55.3 $\pm$ 6.15& 11.8 $\pm$ 0.105\\
            & \multirow[t]{2}*{hex} & 298.15K & 1.71 $\pm$ 0.192& 65.3 $\pm$ 4.89& 12.0 $\pm$ 0.110\\
            &  & 353.15K & 1.42 $\pm$ 0.208& 66.1 $\pm$ 2.58& 11.4 $\pm$ 0.105\\ \midrule
            \multirow[t]{4}{*}{1000}& \multirow[t]{2}*{square}& 298.15K& 0.430 $\pm$ 0.171& 181 $\pm$ 12.0& 10.5 $\pm$ 0.091\\
            & & 353.15K& 0.431 $\pm$ 0.236& 134 $\pm$ 11.6& 11.0 $\pm$ 0.094\\
            & \multirow[t]{2}*{hex}& 298.15K& 1.22 $\pm$ 0.299& 62.9 $\pm$ 4.75& 11.0 $\pm$ 0.100\\
            & & 353.15K& 1.27 $\pm$ 0.291& 97.8 $\pm$ 4.85& 11.0 $\pm$ 0.094\\ \midrule
            \multirow[t]{4}{*}{5000}& \multirow[t]{2}*{square}& 298.15K& 1.04 $\pm$ 0.213& 351 $\pm$ 22.4& 9.45 $\pm$ 0.079\\
            & & 353.15K& 0.408 $\pm$ 0.269& 209 $\pm$ 36.2& 9.70 $\pm$ 0.087\\
            & \multirow[t]{2}*{hex}& 298.15K& 0.949 $\pm$ 0.179& 237 $\pm$ 37.9& 10.0 $\pm$ 0.075\\
            & & 353.15K& 0.381 $\pm$ 0.152& 151 $\pm$ 23.6& 9.47 $\pm$ 0.111\\ \midrule
            \multirow[t]{4}{*}{9000}& \multirow[t]{2}*{square}& 298.15K& 1.60 $\pm$ 0.249& 105 $\pm$ 12.3& 10.3 $\pm$ 0.083\\
            & & 353.15K& 1.21 $\pm$ 0.270& 153 $\pm$ 18.0& 10.3 $\pm$ 0.107\\
            & \multirow[t]{2}*{hex}& 298.15K& 1.64 $\pm$ 0.357& 104 $\pm$ 10.6& 10.2 $\pm$ 0.085\\
            & & 353.15K& 1.11 $\pm$ 0.191& 271 $\pm$ 24.2& 9.55 $\pm$ 0.154\\
            \bottomrule
        \end{tabular}
    \end{table*}

    \begin{figure*}[!ht]
        \centering
        \includegraphics[width=\textwidth]{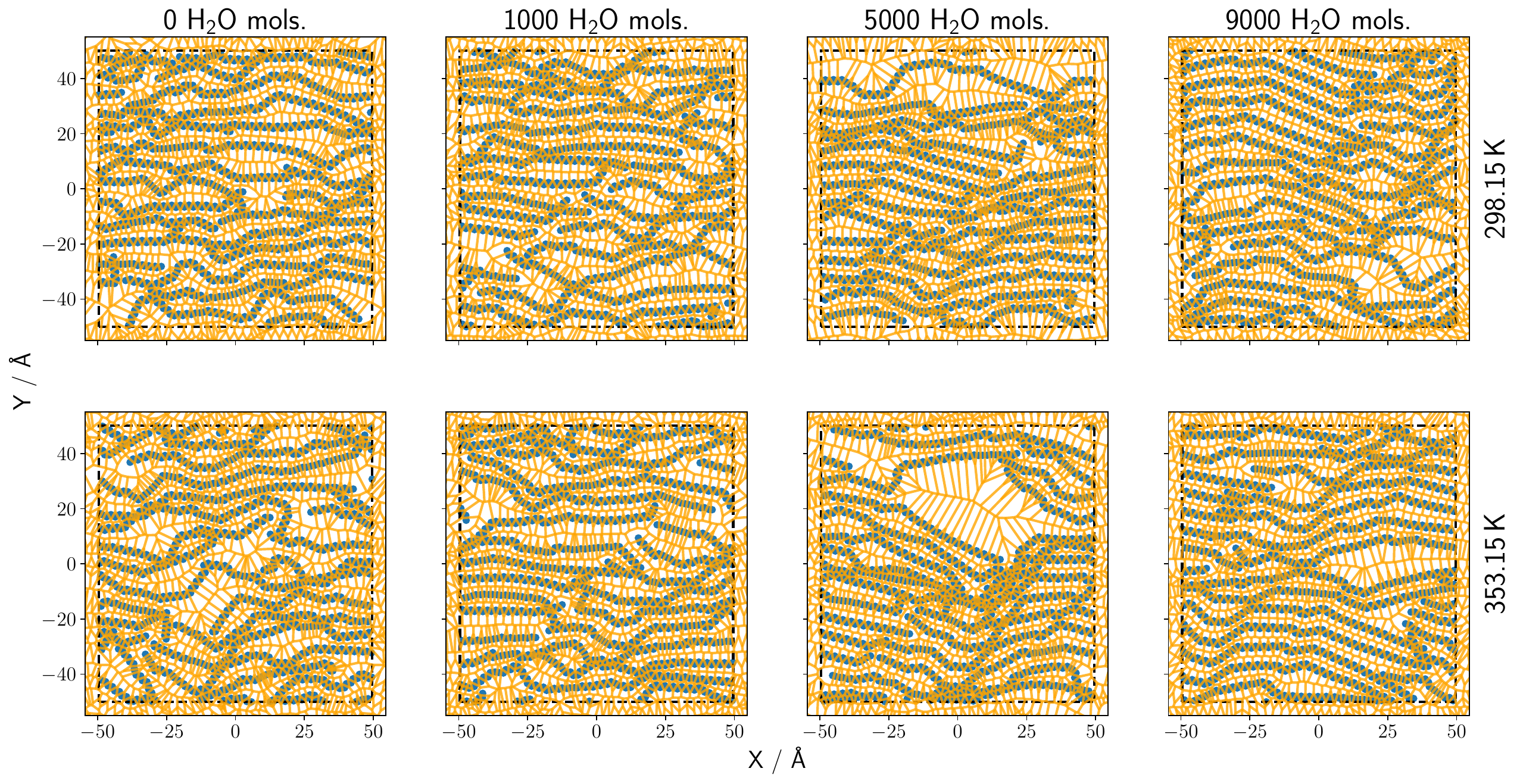}
        \caption{\label{fig:vor_fl_9000}Exemplary Voronoi diagrams for the dense Nafion layers at both investigated temperatures (row wise) and the four different amounts of water molecules added. Blue dots denote carbon backbone atoms, orange lines denote the region edges. Black dashed lines show the simulation cell boundaries.}
    \end{figure*}

\end{document}